\documentclass[11pt,preprint]{article}
\pdfoutput=1 
\usepackage{jcappub, url, enumerate, cancel}
\newcommand\PM[1]{\begin{pmatrix}#1\end{pmatrix}}
\newcommand\g{\gamma}

\title{Warm Dark Matter in Low Scale Left-Right Theory}

\author[a,b]{Miha Nemev\v{s}ek,}
\author[a]{Goran Senjanovi\'c,}  
\author[a]{Yue Zhang\,}
\affiliation[a]{International Centre for Theoretical Physics, Strada Costiera 11, Trieste 34014, Italy}
\affiliation[b]{Jo\v{z}ef Stefan Institute, Jamova 39, Ljubljana 1001, Slovenia}
\emailAdd{miha@ictp.it}
\emailAdd{goran@ictp.it}
\emailAdd{yuezhang@ictp.it}

\abstract{
We investigate the viability of having dark matter in the minimal left-right symmetric theory. We find the lightest right-handed neutrino with a mass around keV as the only viable candidate consistent with a TeV scale of left-right symmetry. In order to account for the correct relic density with such low scales, the thermal overproduction of the dark matter in the early universe is compensated by a sufficient late entropy production due to late decay of heavier right-handed neutrinos. We point out that the presence of the right-handed charge-current interactions, operative around the QCD phase transition, has a crucial impact on the amount of dilution, as does the nature of the phase transition itself. A careful numerical study, employing the Boltzmann equations, reveals the existence of a narrow window for the right-handed gauge boson mass, possibly within the reach of LHC (in disagreement with a previous study). We also elaborate on a variety of astrophysical, cosmological and low energy constraints on this scenario.
}

\keywords{warm dark matter, left-right symmetric model, seesaw mechanism, late entropy production, QCD phase transition}

\arxivnumber{1205.0844}

\begin{document}
\maketitle

%
%
\section{Introduction}

The Left-Right (LR) symmetric theories~\cite{Pati:1974yy, Mohapatra:1974gc, Senjanovic:1975rk, Senjanovic:1978ev} have over the years emerged as one of the main candidates for the theory beyond the Standard Model (SM). Its great achievement is a prediction of non-vanishing neutrino mass, whose smallness gets naturally tied to the maximality of parity violation of weak interactions, through the seesaw mechanism~\cite{Min, MS, Glashow, GRS, yanagida}. This model offers spectacular signatures at colliders such as the Large Hadron Collider (LHC): direct lepton number violation through the production of the heavy right-handed (RH) neutrinos and its subsequent decay, giving a final state with same-sign dileptons~\cite{Keung:1983uu}, the high-energy counterpart of the neutrinoless double beta decay~\cite{Racah:1937qq, Furry:1939qr} and lepton number violating decays of light mesons~\cite{Cvetic:2012hd}. Moreover, the Majorana nature of the RH neutrino manifests itself through the equal production of charged leptons and its anti-particles. A confirmation of the claimed observation of the neutrinoless double beta decay, could even require the LR scale to be tantalizingly close to the LHC reach~\cite{Tello:2010am, Nemevsek:2011aa} (for a review, see~\cite{Senjanovic:2010nq, Senjanovic:2011zz}), if cosmology constraints keep pushing down the sum of neutrino masses~\cite{Fogli:2008ig, Hannestad, Abazajian:2011dt} and thus disfavoring the contribution due to neutrino mass~\cite{Vissani:1999tu}.

There is circumstantial evidence that Dark Matter (DM), which composes about eighty percent of matter energy density in the universe, is in the form of particles. As is well known, the minimal SM fails to provide a DM candidate. At first glance, LR theories could do the job, for they introduce new, potentially stable particles. In particular, the lightest RH neutrino, if light enough, could easily be cosmologically stable, since its Yukawa couplings to the light neutrinos can be arbitrarily small. In a sense, the LR theory can be viewed as a natural framework of DM. Notice there is no need for artificially imposing any $Z_2$ symmetry at high energy -- the $SU(2)_R$ gauge interactions would simply break it. Such gauge interactions cease to be operative in the decay of the lightest RH neutrino, if it is the lightest fermion that couples to $W_R$. An approximate $Z_2$ symmetry emerges at low energy, if all its Yukawa couplings are negligibly small.

A warm DM candidate~\cite{Pagels:1981ke, Peebles:1982ib, Bond:1982uy, Olive:1981ak} with a mass around keV, while it works as well as cold DM for the large scale structure formation, can suppress the structures on smaller scales via free streaming~\cite{Colombi:1995ze}. This scenario is of particular interest as a solution to the problems of very cuspy halo profiles and over-populated low-mass satellite galaxies, usually predicted by cold DM. The idea of having RH neutrino as warm DM candidate with a mass around a keV was introduced around thirty years ago~\cite{Olive:1981ak, Dodelson:1993je}. Due to the presence of gauge interactions, one expects the RH neutrino playing the role of DM to have a similar relic number density as the one of the light neutrinos, if the scale of LR symmetry is not far above the electroweak scale. In~\cite{Olive:1981ak}, the first cosmological bounds on stable heavy neutrinos, charged under a new gauge symmetry, were studied and the problem of their potential over-abundance was stressed. Ref.~\cite{Scherrer:1984fd} offered a nice way out, by today a text-book scenario~\cite{Kolb:1990vq1}, through the late entropy injection due to the decay of a heavier long-lived particle, for example the heavier RH neutrino~\cite{Asaka:2006ek}.  

Using the idea of~\cite{Asaka:2006ek}, a few years ago Bezrukov et al~\cite{Bezrukov:2009th}  performed studied this issue in the context of LR theories and argued that one cannot obtain the correct DM abundance unless the mass of the RH charged gauge boson $W_R$ is above 10\,--\,16 TeV, far from the LHC reach. This unfortunate result made it irresistible for us to reconsider their analysis with great care. While we agree with the basic mechanism presented in~\cite{Bezrukov:2009th}, our analysis reveals an additional window for the $W_R$ mass around roughly 5 TeV, possibly within the LHC reach. This is the main result of our paper, whose importance cannot be over-emphasized. 

The key point in our work of realizing such low scale LR symmetry is to take advantage of the QCD phase transition, where the number of relativistic degrees of freedom changes dramatically. Depending on the flavor structure of the  their gauge couplings,  the RH neutrinos decouple at different temperatures, which could be separated enough to lie before and after the transition. We show that this plays a crucial role in producing large enough amount of entropy, in order to dilute the DM relic abundance towards the acceptable range.

In the decoupling limit, when the LR scale is large, our picture crosses over smoothly to the so-called $\nu$MSM scenario, studied extensively over the years~\cite{Asaka:2005an, Asaka:2006ek, Kusenko:2009up} (for a review on the topic of light sterile neutrinos, see~\cite{Abazajian:2012ys} and references therein). In the $\nu$MSM case, one has only the SM augmented with RH neutrinos and Yukawa interactions take over the role of the RH gauge interactions, leading to the sterile neutrino picture. This transition is quantified carefully in Sec.~\ref{secPhaseDiagram}. In other words, the LR theory cannot fail to account for dark matter, if its scale is high enough. However, the theory then stops being directly verifiable and loses most of its phenomenological appeal. This is why we find the existence of the narrow band for not-so-heavy $W_R$ important enough to warrant another paper on the subject.


In order to ease our reader's pain, we have decided to make this presentation as pedagogical as possible and therefore we will start from scratch in presenting our work. In the following Sec.~\ref{secLRModel}, we will review the essential features of the LR symmetric model and summarize the present-day theoretical and experimental status. Section~\ref{secTwoNs} presents all the central ideas behind this works in an accessible way, using rough estimates, which paves the way for Sec.~\ref{secN2N3Delta}, where a detailed numerical analysis is carried out. In Sec.~\ref{secConstraints} we discuss additional astrophysical and cosmological constraints, together with the limit coming from the search for neutrinoless double beta decay. We conclude in Sec.~\ref{secConclusions}.

%
%
\section{Minimal Left-Right Model: A Telegraphic Review \label{secLRModel}}

The left-right symmetric model is based on the $SU(2)_L \times SU(2)_R \times U(1)_{B-L}$ gauge group (suppressing color), supplemented by a symmetry between the left and the right sector~\cite{Pati:1974yy, Mohapatra:1974gc, Senjanovic:1975rk, Senjanovic:1978ev}. Quarks and leptons come in symmetric representations
\begin{equation}
	Q_{L,R} = \begin{pmatrix} u \\ d \end{pmatrix}_{L,R}, \quad L_{L,R} = \begin{pmatrix} \nu \\ \ell \end{pmatrix}_{L,R}.
\end{equation}
The Higgs sector of the minimal model~\cite{Min, MS} consists of a bidoublet $\Phi = \left( 2_L, 2_R, 0_{B-L} \right)$ and two triplets, $\Delta_L = \left(3_L, 1_R, 2_{B-L} \right)$ and $\Delta_R = \left(1_L, 3_R, 2_{B-L} \right)$
\begin{equation}  \label{ds32}
  \Phi = 
  \begin{pmatrix} \phi_1^0 & \phi_2^ + 
  \\
  \phi_1^- & \phi_2^0 \end{pmatrix}\, ,
  \quad
  \Delta_{L, R} = 
  \begin{pmatrix} \Delta^+ /\sqrt{2} & \Delta^{++} 
  \\
  \Delta^0 & -\Delta^{+}/\sqrt{2}
  \end{pmatrix}_{L,R} \, .
\end{equation}
Hereafter, we refer to this setup as the minimal LR standard model (LRSM). The symmetry breaking in the model~\cite{Mohapatra:1980yp} is characterized by the following vacuum expectation values (vev) $\langle \Phi \rangle = \text{diag}\left( v_1, v_2 \right)$, $\langle \Delta^0_{L,R}\rangle = v_{L,R}$, which have a hierarchical order $v_L^2 \ll v^2 = v_1^2 + v_2^2 \ll v_R^2$, with $v = 245 \text{ GeV}$. For a recent summary of the main features, see~\cite{Zhang:2007da, Maiezza:2010ic, Blanke:2011ry}. The resulting masses of the heavy gauge bosons are $M_{W_R} = g \, v_R$ and 
\begin{equation}
  \quad M_{Z_{LR}} \simeq \sqrt{3} \, M_{W_R}.
\end{equation}
Notice that $Z_{LR}$ is appreciably heavier than $W_R$, which turns out to be important in the study of the RH neutrino freeze-out.  The vev's of the bi-doublet give masses to the charged fermions and Dirac masses to neutrinos. The triplet vev $v_R$ gives directly Majorana mass to RH neutrinos, which results in the type-I seesaw mechanism~\cite{Min, MS, Glashow, GRS, yanagida}, while $v_L$ independently gives Majorana light neutrino masses, as in the so-called type-II seesaw~\cite{Magg:1980ut, Mohapatra:1980yp, Lazarides:1980nt}. As we  show in what follows,  the seesaw picture is a must for the DM scenario to work.

The charged and neutral currents relevant for our study, up to tiny $v^2 / v_R^2 \lesssim 10^{-3}$ corrections (see below for the limits on the LR scale), are
\begin{eqnarray}
 Ê{\mathcal L}_{CC} &=& \frac{g}{\sqrt{2}} W_R^\mu \left[
  \begin{pmatrix} \overline{N_1} &Ê\overline{N_2} &Ê\overline{N_3} \end{pmatrix}_R
  \mathbf{V}_\ell^{R\dag} \g_\mu \PM {e \\ \mu \\ \tau}_{\!\!R} + \begin{pmatrix} 
 Ê\overline{u} &Ê\overline{c} &Ê\overline{t} \end{pmatrix}_R
  \mathbf{V}_q^R \g_\mu \PM {d \\ s \\ b}_{\!\!R} \right]+\text{h.c.} \,, 
  \\
  {\mathcal L}_{NC} &=& \frac{g}{\sqrt{1 - \tan^2 \theta_W}} Z^\mu_{LR} \bar f \gamma_\mu \left[ T_{3R} + \tan^2\theta_W (T_{3L} - Q) \right]f + \frac{g m_N}{2M_{W_R}} \Delta_R^0 N N \label{neutralcurrent}
\end{eqnarray}
where $N$'s are defined as RH neutrinos mass eigenstate states, and we have taken $g_R = g_L\equiv g$, appropriately for a LR symmetric theory. The mixing matrices $\mathbf{V}_{\ell}^R$ and $\mathbf{V}_q^R$ are the right-handed analogues of the left-handed PMNS and the CKM mixing matrices with elements $V^R_{\ell N}$ and $V^R_{qq'}$. Of course, there is no reason that $g_R = g_L$ relation should hold exactly at the scale of interest; there could be easily a variation if the LR symmetry is broken at a high scale. The small effects due to renormalization group equation running would change none of our conclusions.

Notice that $Z_{LR}$, besides being heavier than $W_R$, has also smaller couplings to $N$'s. Thus, a RH neutrino that couples only to $Z_{LR}$ will decouple earlier in the thermal history of the universe for a given LR scale. Since a warm DM candidate in this kind of a setup is typically overproduced, it will turn out desirable to profit from this fact and decouple it from the $W_R$ at relevant temperatures.

\begin{table}
\centering
\begin{tabular}{|c|c|r|c|c|}
	\hline
	Particle & Final state & Lower limit $   $ & Collaboration & Comments
	\\ \hline \hline
	$W_R$ & $jj$ & 1.5 TeV 	& CMS~\cite{CMSjj} & independent on $N$ mass
	\\
	$W_R$ & $e/\mu + N$ &	2.5	TeV 	& CMS~\cite{CMSWR} & light $N$ (missing energy)
	\\
	$W_R$ & $\ell \ell j j$  & $\lesssim 2.5$ TeV	& ATLAS, CMS~\cite{ATLASKS, CMSKS} & heavy Majorana $N$~\cite{Nemevsek:2011hz}
	\\
	$Z_{LR}$ & $e^+ e^-/\mu^+ \mu^-$ & $\sim 2$ TeV  & ATLAS~\cite{ATLASZLR} & see~\cite{Langacker:2009su}
	\\
	$Z_{LR}$ & $e^+ e^-$ & $\sim 3$ TeV  & LEP~\cite{LEPZLR} & indirect, see~\cite{Carena:2004xs, Cacciapaglia:2006pk}
	\\
	$\Delta_L^{++}$ & $\ell_i^+\ell_j^+$ & 100-355 GeV & ATLAS~\cite{Aad:2012cg} & spectrum dependent~\cite{Melfo:2011nx}
	\\
	$\Delta_L^{+}$ &  $\cancel{E}_T+j$ & 70-90 GeV & LEP~\cite{Abbiendi:2003ji} & chargino search~\cite{Pierce:2007ut}
	\\	
	$\Delta_L^{0}$ &   & 45 GeV & LEP~\cite{LEPZLR} & $Z$-boson width
	\\	
	$\Delta_R^{++}$ &$\ell_i^+\ell_j^+$ & 113-251 GeV & ATLAS~\cite{Aad:2012cg}, CDF
	\cite{Abazov:2011xx} & flavor dependent
	\\
	\hline
\end{tabular}
\caption{A summary of limits on the mass scales of the particles in LRSM from collider searches. 
\label{tabBound}}
\end{table}

The most stringent theoretical limit on the LR scale is derived from neutral kaon mixing~\cite{Beall:1981ze, Mohapatra:1983ae}, and the latest studies set a lower bound $M_{W_R} > 2.5\,$--\,$4 \text{ TeV}$~\cite{Maiezza:2010ic, Zhang:2007fn, Zhang:2007da}, depending on the choice of the LR symmetry, charge conjugation or parity, respectively. The experiment, however, is now catching up and the theoretical constraints are becoming obsolete. Direct searches are continuously pushing up the limits on mass scales in the LRSM and we summarize them in Table~\ref{tabBound}. The window around $\sim 5 \text{ TeV}$ that will emerge from our DM study is comfortably above all the current theoretical and experimental bounds. It is also noteworthy that the second Higgs doublet belonging to $\Phi$, orthogonal to the SM-like one, must be heavier than $\sim$\,10\,TeV due to the contribution to tree-level flavor changing processes.  For a recent complete study of a variety of flavor processes in the LRSM, see~\cite{Blanke:2011ry}.

Moreover, Table~\ref{tabBound} tells us that most of the states have masses around or above the weak scale. The
only exception are the RH neutrinos $N$ and the neutral Higgs $\Delta_R^0$, which behave like singlets under the SM gauge group and are not very much constrained by collider searches. 
They may be long-lived and are therefore potential DM candidates. In the next section we study which, if any, can 
actually do the job.

%
%
\section{A Tale of Three Right-handed Neutrinos \label{secTwoNs}}

We discuss here the history and role of RH neutrino as DM in the early universe, in the context of TeV scale LRSM. It contains the essence of what is going on: the lightest $N$ is presumably the DM due to its longevity, while the heavier one(s) should make sure that its abundance is correctly accounted for.

This section is the core of our work; it is here that our reader will find the central ideas, albeit simplified. Thus, we urge her to postpone the coffee break until having gone through it. The technicalities required for a precise quantitative picture are left for the Sec.~\ref{secN2N3Delta}.

\subsection{Warm Dark Matter Candidate \label{candidates}}

Let us now go through the list of potential candidates for the dark matter in the LRSM. From the discussion in the previous section, it is
clear that the neutral components belonging to the Higgs bi-doublet $\Phi$ and the triplet $\Delta_L$ are heavy enough to decay on collider time scales, and are thus ruled out from the start.

\begin{itemize}
  
\item One possibility is the neutral component of the $SU(2)_R$ triplet, ${\rm Re}\Delta^0_R$. Being a SM singlet, it is 
allowed to be as light as one wishes, and if it is lighter then the RH neutrinos, it will decay into two photons, with a rate
\begin{equation}\label{deltalife}
  \Gamma_{\Delta^0_R\to \gamma\gamma} \simeq \frac{49}{8 \pi} \left( \frac{\alpha}{4\pi} \right)^2 
  \left( \frac{M_{W}}{M_{W_R}} \right)^2  \frac{G_F}{\sqrt{2}} m^3_{\Delta} \simeq 10^{-50} \text{ GeV}
  \left(\frac{m_{\Delta}}{\rm keV} \right)^3 \left( \frac{10^{12} \text{ GeV}}{M_{W_R}} \right)^{2},
\end{equation}
where $49$ is the loop function squared. This approximate formula takes into account only the dominant contribution due to the heavy charged $W_R$. This vector boson dominance resembles the situation in the SM, where the gauge contribution is roughly an order of magnitude bigger than the rest, i.e. the fermionic one. Here, there is no fermionic contribution, but instead the one from the charged scalars, which is typically much less than the gauge boson one.

There is a stringent lower limit on the stability of a radiatively decaying dark matter particle~\cite{Abazajian:2001vt}, $\tau_\Delta \gtrsim 10^{26}\, \text{sec}$, or equivalently $\Gamma \lesssim 10^{-50}\,\text{ GeV}$, up to uncertainties in astrophysical parameters. This pushes the scale of LR symmetry far away from the LHC hope into the despair of no direct detection.  
It is somewhat  surprising though, that the $\Delta_R^0$ could be a viable dark matter candidate in an $SO(10)$ grand unified theory, where the LR symmetry breaking scale lies preferably around $10^{10} - 10^{12} \text{ GeV}$~\cite{Rizzo:1981su, Bertolini} (admittedly it would have to be incredibly light, creating yet another hierarchy problem). In any case, this possibility would take us far away from our search for a low scale LR symmetry. 

\item Thus, for low LR scale, the only viable candidate left is the lightest RH neutrino, to which for definiteness we refer as $N_1$. For its mass below the pion mass, the decay channel mediated by $W_R$ closes quickly. It can only be destabilized by Dirac Yukawa couplings due to the mixing with left-handed neutrinos. This mixing leads to its decay to a light left-handed neutrino and a monochromatic photon, or three light neutrinos. Again, the X-ray constraints implies such mixing to be tiny~\cite{Smirnov:2006bu, Boyarsky:2009ix}
\begin{equation}\label{theta}
  \theta_1^2 < \left(1.8-3.1\right)\times 10^{-5} \left( \frac{1\,\rm keV}{m_{N_1}} \right)^{5}.
\end{equation}
\end{itemize}

In the rest of this work, we thus study the exciting possibility of $N_1$ playing the role of dark matter. The most reliable cosmological lower limit on the DM mass is derived by considering the phase space density of compact objects, which is around $\sim$\,keV scale~\cite{Tremaine:1979we, Boyarsky:2008ju, Gorbunov:2008ka}. For $M_{W_R}$ lying in $1-10\text{ TeV}$ region, one may worry whether such small mass of $N_1$ is consistent with the seesaw formula, with radiative corrections included. The real question is the radiative stability of the neutrino Dirac mass, studied in Ref.~\cite{Branco:1978bz}, with a conclusion that it can be as small as a few eV.  This in turn implies (via the seesaw) that the RH neutrino is allowed to be as light as $10-100\text{ eV}$. In other words, a keV RH neutrino is a perfectly natural choice for a DM candidate.

The main obstacle we have to face when $M_{W_R}$ lies in the TeV region, turns out to be the over-abundance of $N_1$~\cite{Olive:1981ak, Bezrukov:2009th}, because the $SU(2)_R$ gauge interactions keep it in thermal equilibrium when the temperature is high. Intuitively, one expects $N_1$ to decouple at a temperature not far above the usual decoupling temperature of light neutrinos, so that its density is also similar to that of light neutrinos -- a disaster for a particle with a mass above keV. Obviously, the heavier it is, the more severe a problem this becomes. The bottom line is that one is pushed to the picture~\cite{Dodelson:1993je} of warm dark matter, with a mass in the keV range. This is welcome due to the need to suppress the small-scale structures. 

Furthermore, it has been pointed out that supernovae cooling imposes a tight constraint on the flavor structure of weakly interacting species lighter than 10\,MeV~\cite{Raffelt:1987yt}. In the context of LRSM, this implies $M_{W_R}>\sqrt{|V^R_{e1}|} \times 23\,$TeV~\cite{Barbieri:1988av}, which means that the electron component of $N_1$ in the RH charged current has to be roughly below 1\%, if LR symmetry is close to the TeV scale.

In what follows, we shall stick to the keV warm dark matter RH neutrino and pursue its implications to the bittersweet end.

\subsection{Thermal production via freeze out}

The presence of new gauge interactions in the LRSM has a major impact on the thermal production of dark matter in the form of RH neutrinos. If the universe starts from a sufficiently high temperature, these interactions will keep them in thermal equilibrium via scatterings with the SM fermions. On the other hand, it turns out that the Dirac Yukawa couplings of a keV RH neutrino are never large enough to bring it into equilibrium above the electroweak scale. Also, after the electroweak symmetry breaking, the mixing between $N_1$ and light neutrinos in matter are suppressed by the finite density potential~\cite{Barbieri:1990vx, Enqvist:1990ad}, compared to the vacuum mixing angle $\theta_1$. Therefore, the SM weak interactions cannot bring RH neutrinos in equilibrium~\cite{Boyarsky:2009ix}. Although the direct thermal production via the mixing is negligible, non-thermal contributions due to accumulative oscillations~\cite{Dodelson:1993je, Shi:1998km, Abazajian:2001nj, Boyanovsky:2006it} may become significant, as in the case of $\nu$MSM. However, this effect is sub-dominant compared to the thermal (over-)production in the LRSM as discussed below.

Let us quantify more precisely how the dark matter RH neutrino $N_1$ with TeV scale gauge interactions typically gets over-produced in the usual thermal history. The freeze-out temperature $T_f$ can be estimated by the out-of-equilibrium condition $\Gamma = H$, where $\Gamma$ is the annihilation rate of $N_1$ and $H$ is the Hubble parameter. The interactions that keep them in thermal equilibrium are scatterings of $N$ with the light SM fermions, mediated by the heavy gauge bosons, $W_R$ and $Z_{LR}$ in the LRSM. In the radiation dominated era, we have~\cite{Kolb:1990vq2}
\begin{equation}
	G_F^2 \left( \frac{M_W}{M_{W_R}} \right)^4 T_f^5 \simeq  \sqrt{g_*(T_f)} \frac{T_f^2}{M_{\rm p}} \, .
\end{equation}
Generally, for any RH neutrino that decouples from equilibrium while still relativistic, one can obtain the freeze-out temperature $T_f$ as a function of $M_{W_R}$ and the number of degrees of freedom $g_*(T_f)$ at that time. 

For the $W_R$ in the TeV region, the freeze-out temperature is around
\begin{equation} \label{eqTfEstimate}
	T_{f} \simeq 400\, {\rm MeV} \left(\frac{g_{*}(T_f)}{70}\right)^{1/6} \left( \frac{M_{W_R}}{5\,{\rm TeV}} \right)^{4/3} \ .
\end{equation}
where the normalization reflects the fact $g_*(400\, \text {MeV}) \simeq 70$.
For a RH neutrino $N$, which freezes out while relativistic, the number per entropy density is approximately
\begin{equation}\label{Nyield}
Y_{N} \equiv \frac{n_{N}}{s} \simeq \frac{135 \, \zeta(3)}{4 \pi^4 \, g_*(T_{f})} \ .
\end{equation}
The yield $Y_N$, being thermally conserved quantity, turns out to be useful in what follows. It may be suppressed if $T_f$ is higher, which can happen if neutral current interactions play the dominant role in the freeze-out, as pointed out above. Such a situation occurs, when the presence of the charged lepton corresponding to a RH neutrino  in the thermal bath is Boltzmann suppressed. With the above values of the freeze-out temperature, the optimal option is to couple the DM candidate $N_1$ predominantly to $\tau$. 
This is the first crucial ingredient towards determining the flavor structure in the RH charged currents necessary to accommodate DM in this theory. Remarkably enough, when the dust settles we will end up with a completely determined flavor structure of the charged current.

In any case, a keV $N_1$ is still relativistic at such temperatures, therefore its relic abundance today is roughly
\begin{equation} \label{dmrelic}
  \Omega_{N_1} = \frac{Y_{N_1} m_{N_1} s}{\rho_c} \simeq 3.3\times \left( \frac{m_{N_1}}{1\,{\rm keV}} \right) \left( \frac{70}{g_{*}(T_{f1})} \right) \ ,
\end{equation}
where we have used today's entropy density $s=2889.2\,{\rm cm}^{-3}$, critical density $\rho_c=1.05368\times10^{-5} h^2\,{\rm GeV/cm^3}$ and $h=0.7$. This is to be contrasted with the observed dark matter relic abundance of the universe~\cite{Komatsu:2008hk}
\begin{equation}\label{pdgrelic}
  \Omega_{\text{DM}} = 0.228 \pm 0.039 \ , 
\end{equation}
at $3\sigma$ confidence level. Clearly, the estimated contribution of $N_1$ in Eq.~\eqref{dmrelic} over-closes the universe, when $W_R$ lies in the TeV region, by at least a factor of $\sim 12.5 \times ({m_{N_1}}/{1\,{\rm keV}})$. Since there is no room in the minimal LRSM for $\sim$1000 degrees of freedom in $g_*(T_{f1})$, this problem cannot be solved by simply raising the scale $M_{W_R}$. Such a possibility of a huge number of ad-hoc new states is rather unappealing in any case.
 
\subsection{Late entropy production}

The only way out of this impasse is to dilute the number density of $N_1$ by entropy production due to the late decay of some massive particle which dominates the universe~\cite{Scherrer:1984fd}. Such a late decay should inject relativistic light SM particles that quickly equilibrate with the thermal plasma and ``reheat" the photon temperature. In turn, it takes longer for the photons to cool down to present-day temperature and the number density of DM is effectively reduced. In order for the dilution to work, the temperature of $N_1$ should not increase, and it is therefore crucial that $N_1$ itself is not a decay product of the heavy decaying particle.

In order to release a substantial amount of entropy, such a particle is required to be long-lived, with lifetimes up to a second. In the minimal LRSM, the only particles relevant for the late decay are the following.
\begin{itemize}
\item As discussed at the beginning of Sec.~\ref{candidates}, the neutral component of the $SU(2)_R$ triplet ${\rm Re}\Delta^0_R$ can be as light as one wishes. For TeV scale LR symmetry, it has to be lighter than MeV in order to live as long as one second. This makes it too light to play any significant role in entropy production.
\item The heavier RH neutrinos $N_{2,3}$ are the only remaining viable candidates and they play the role of diluters in this 
scenario~\cite{Asaka:2006ek}. This section focuses on determining their characteristics, required for adequate relic abundance of DM. Here, we refer to the diluter as a generic $N$, and use explicit indices ($2, 3$) when necessary to specify the flavor structure. The DM RH neutrino is always $N_1$.
\end{itemize}
In order to achieve a sufficient dilution, the mass of the diluter $m_N$ should not exceed its freeze-out temperature $T_f$, otherwise the yield in Eq.~\eqref{Nyield} receives an additional Boltzmann suppression factor $e^{-m_N / T_f}$. If the gauge interactions of RH neutrinos are universal, their freeze-out temperatures is of the same order (we return to this point in Sec.~\ref{secwindow}) and similar yields for all RH neutrinos $Y_{N_i}$ are expected. 

As the temperature of the universe drops, a sufficiently massive and long-lived RH neutrino can temporarily dominate the total energy density. After $N$ decays, the energy density is transferred into that of radiation. In the sudden decay approximation, all the $N$'s decay at $t \simeq \tau_{N}$ and ``reheat" the universe to the temperature $T_r$~\cite{Scherrer:1984fd}
\begin{equation}
  T_{r} \simeq 0.78 \, g_*(T_{r})^{-1/4} \sqrt{\Gamma_N M_{\rm p}} \simeq 1.22\,{\rm MeV} \left( \frac{1\,{\rm sec}}{\tau_N} \right)^{1/2} \ .
\end{equation}
In order to start the Big Bang Nucleosynthesis (BBN) with a correct proton-neutron number ratio, $T_{r}$ should be larger than about MeV, which gives an upper bound on diluters' lifetime $\tau_{N}\lesssim \mathcal{O}(1)\,$second.

Using energy conservation $m_N n_{N}(\tau_{N}) \equiv m_N Y_N s =\rho_{R}(T_r)$, or in the other words

\begin{equation}\label{goran_wants_to_label_it}
  m_N  Y_N s_{before} = \frac{3}{4} s_{after}T_r\ ,  
 \end{equation}
the dilution factor, defined as the ratio of entropy before and after the decay, in the sudden decay approximation (no volume change) becomes roughly
\begin{equation}\label{eqDilutionS}
  \mathcal{S} \equiv \frac{S_\text{after}}{S_{\text{before}}} \simeq  \frac{s_\text{after}}{s_{\text{before}}}\simeq 1.8 \left(g_*(T_r)\right)^{1/4} \frac{Y_{N} \, m_{N}}{\sqrt{\Gamma_{N} M_{\rm p}}} \ ,
\end{equation}
where $\Gamma_{N}=\tau_{N}^{-1}$ and $M_{\rm p} = 1.2 \times 10^{19} \, $GeV is the Planck scale. If such entropy production happens well after $N_1$ froze-out, the relic density calculated in Eq.~(\ref{dmrelic}) will be reduced by the dilution factor, $\Omega_{N_1} \to \hat\Omega_{N_1}= \Omega_{N_1}/\mathcal{S}$. For a reheating temperature $T_r$ around MeV, 
\begin{equation} \label{dilutedmrelic}
  \hat \Omega_{N_1} \simeq (0.228 + 0.039) \left( \frac{m_{N_1}}{1\,{\rm keV}} \right) \left( \frac{1.85\,{\rm GeV}}{m_N} \right) \left( \frac{1\,{\rm sec}}{\tau_N} \right)^{1/2} \left( \frac{g_{*}(T_{f2,3})}{g_{*}(T_{f1})} \right) \ .
\end{equation}

The dilution factor in Eq.~\eqref{eqDilutionS} is proportional to the lifetime of the diluter $N$, which should be long enough to make a sufficient impact on the relic density of $N_1$. On the other hand, a late decaying particle with a lifetime longer than about a second may threaten the success of BBN. Therefore, the optimized situation is to have $\tau_{N}\sim1\,$sec (or $T_r \sim$\,MeV, as assumed above) which, depending on the scale of LR symmetry, narrows down the mass range of the diluter $N$, since the same gauge interactions that govern the freeze-out of RH neutrinos are responsible for the decay of $N$.

Let's recapitulate once again the logic of our search for the light $W_R$ accessible to experiment. This immediately narrows down the freeze-out temperature of $N$'s in the few 100 MeV range, as seen in Eq.~\eqref{eqTfEstimate}. In order to avoid the Boltzmann suppression, the diluters mass should be below $T_f$, which in turn brings back the problem of over-abundance of DM, since the third term in Eq.~\eqref{dilutedmrelic} becomes large. We are back to square one, it seems. However, there is a potential way out; make the last term small by separating the freeze-out temperatures of the diluters and the dilutee. It turns out that the nature of the QCD phase transition plays an essential role in this, as discussed in the coming section. 

A lifetime of around one second restricts the mass of $N$ to a narrow region in the few 100 MeV range for a $W_R$ in the few to ten TeV region. Depending on its mass, $N$ decays via heavy $W_R$ either predominantly mainly into a lepton plus two light quarks (which later hadronize), with a lifetime
\begin{equation} \label{eqNljj}
  \tau(N_i\to\ell jj) = \frac{192 \pi^3}{G_F^2} \left(\frac{M_{W_R}}{M_W} \right)^4 \frac{1}{|V_{ud}^R V^R_{\ell i}|^2 m_N^5}
  = 1\,{\rm sec} \left(\frac{2\,{\rm GeV}}{m_N} \right)^{5} \left( \frac{M_{W_R}}{100\,{\rm TeV}} \right)^{4} \ ,
\end{equation} 
or (if $m_N \gtrsim m_{\pi}+m_\ell$, i.e., near the threshold) into a lepton and a pion, with a lifetime~\cite{Nemevsek:2011aa}
\begin{equation} \label{eqNlPi}
\begin{split} 
  \tau(N_i\to\ell\pi) &= \frac{8 \pi}{G_F^2} \left(\frac{M_{W_R}}{M_W} \right)^4
  \frac{1}{|V_{ud}^R V^R_{\ell i}|^2 f_\pi^2 m_N^3} \frac{1}{f(x_\ell, x_\pi)}
  \\
  &= 1 \text{ sec} \left(\frac{m_{N}}{250\,{\rm MeV}}\right)^{-3} \left( \frac{M_{W_R}}{5\,{\rm TeV}} \right)^{4} \left( \frac{0.002}{f(x_\ell, x_\pi)} \right)\ ,
\end{split}
\end{equation}
where $f(x_\ell, x_\pi) =\left[ (1-x_\ell^2)^2 - x_\pi^2(1+x_\ell^2) \right] \left[ \left( 1-(x_\pi+x_\ell)^2 \right)\left( 1-(x_\pi-x_\ell)^2 \right) \right]^{1/2}$ and $x_{\pi, \ell}=m_{\pi, \ell}/m_N$~\footnote{Due to a small mixing angle $\theta^2_{N} \simeq m_\nu/m_{N}$, the $N \to \pi^0 \nu$ channel turns out to be subdominant.}. In the above estimates, we took $V_{ud}^R \simeq V_{ud}^{\rm CKM} \approx 1$. From the above expressions for $N$ lifetime, it is clear that the best bet to have $M_{W_R}$ in the TeV region is to have the pionic decay dominant, together with a final-state phase space suppression. This narrows down $m_{N}$ to lie around 
\begin{equation} \label{eqmN2Pi}
  m_{N} \approx m_\pi + m_\ell \ ,
\end{equation}
together with a lepton mixing $V^R_{\ell\, 2} \simeq V^R_{\ell\, 3} \approx 1$, up to  $\sim1\%$ (see Fig.~\ref{figPhaseSpace}).

\subsection{Fixing the flavor structure}

Let us take a closer look at the leptonic flavor of the diluting RH neutrino $N$ in Eq.~\eqref{eqmN2Pi}. In order to successfully dilute the DM relic density in Eq.~\eqref{dilutedmrelic}, one would naively conclude that $\ell = \tau$ is favored, since the mass of the diluter is in the right ballpark. However, there are several serious drawbacks related to this channel.

First and foremost, in order to avoid the non-relativistic Boltzmann suppression, the freeze-out temperature of $N$ should be bigger than its mass. In order to lift this suppression a high freeze-out temperature is needed, which which for $m_N$ around 1.85 GeV,
requires the $W_R$ boson mass to be heavier than about 15 TeV~\cite{Bezrukov:2009th}, see~Eq.~\eqref{eqTfEstimate} (see also Fig.~\ref{figPhaseSpace}). Being far out of the LHC reach, this is outside of the region of our interest. This by itself prevents any diluters to couple predominantly to the tau.

There is yet another reason that strengthens this result. Namely, the decaying $N$ might have an appreciable branching ratio to $N_1$, which makes the task of dilution more challenging. 
Namely, the cosmological lower limit on the warm DM mass depends on its free-streaming length, which is proportional to the average energy $\langle p_{N_1} \rangle$ at injection~\cite{Bond:1980ha}:
$\lambda_{fs} \simeq 1 {\rm Mpc} \left( \langle p_{N_1} \rangle / \langle p_\nu \rangle\rule{0mm}{3.5mm} \right) \left( {\rm keV}/{m_{N_1}} \right) \mathcal{S}^{-1/3}$.
%
%
The analysis of the Lyman-$\alpha$ forest demands the DM free-streaming length not much longer than a Mpc. Unless the mass of $N$ happens to lie within few MeV above the production threshold, the $N_1$ in the decay product is always too energetic. The free streaming of $N_1$ will erase any structure at scales larger than a Mpc, unless $m_{N_1}$ is much bigger than keV. From Eq.~\eqref{dilutedmrelic} such a heavy mass is disastrous for relic density, for a TeV LR scale.

Therefore, in what follows, we focus either on the electron or the muon flavor channel for the diluters. In this case, the decays of $N_{2,3}$ to (an energetic) $N_1$ are kinematically forbidden if $N_1$ couples predominantly to $\tau$, which is the preferred freeze-out scenario anyway. Namely, in that case the freeze-out temperature of $N_1$ is dictated by the new neutral currents only, suppressed compared to
the charged ones, which in turn makes $N_1$ decouple before the diluters.
At this point, a clear picture of both the flavor structure and the mass spectra of RH neutrinos emerges
\begin{align}\label{flavorspectra}
	\mathbf{V}^R_\ell &\approx \begin{pmatrix}
		0 & 0 & 1
		\\
		0 & 1 & 0 
		\\
		1 & 0 & 0 
	 \end{pmatrix}, \quad \quad
	 \begin{array}{rl}
	 m_{N_1} & \sim \text{ keV},
	 \\
	 m_{N_2} & \approx m_\pi + m_\mu,
	 \\
	 m_{N_3} & \approx m_\pi + m_e.
	 \end{array}
\end{align}
Notice that any departure from diagonallity in the $2-3$ sector effectively reduces the masses of diluters and therefore decreases the dilution factor. In other words, we end up with a flavor-diagonal RH leptonic mixing matrix, to which we stick for the rest of this paper. 

Such a flavor structure is completely different from the PMNS mixing matrix in the left-handed sector. In the case of the type II seesaw, due to the LR symmetry, the flavor structure in the left and right sectors has to be identical. This means clearly that the contribution from the type II seesaw can only play a subdominant role. Let us see now how the type-I seesaw dominance for neutrino masses ends up being consistent with the diluters' lifetimes.
In fact, the partial lifetime of a RH neutrino induced by the Dirac yukawa coupling in the type I case is
\begin{equation}
  \tau_N = \left( \frac{m_\mu}{m_N} \right)^5 \left( \frac{m_D}{m_N} \right)^2 \tau_\mu \simeq \left( \frac{m_\mu}{m_N} \right)^4 \frac{m_\nu}{m_\mu} \, \tau_\mu \simeq \left(\frac{0.1\text{ GeV}}{m_N}\right)^4 10^{2} \text{ sec}.
\end{equation}
For the diluters with mass around 0.1\,GeV, this is certainly a subdominant channel in the presence of TeV gauge interactions and they can fully participate in the seesaw. This is in contrast to the case of $m_{N} \gtrsim \text{ GeV}$, where the Dirac mass has to be suppressed to ensure enough dilution, in which case one would end up with type II dominance (as in Ref.~\cite{Bezrukov:2009th}). Whereas in the latter case, the $W_R$ is necessarily heavy, far above the LHC reach, the appealing aspect of the former case is an emergence of a light $W_R$ window, on which we focus.

The bottom line then is, we end up with fairly light RH neutrinos and negligible leptonic mixing, which eliminates completely both the lepton number violation at colliders and lepton flavor violation. A positive future result in any such process would kill the picture envisioned here, and dark matter would have to come from elsewhere.  

\begin{figure}[t!]
\centerline{\includegraphics[width=8.5cm]{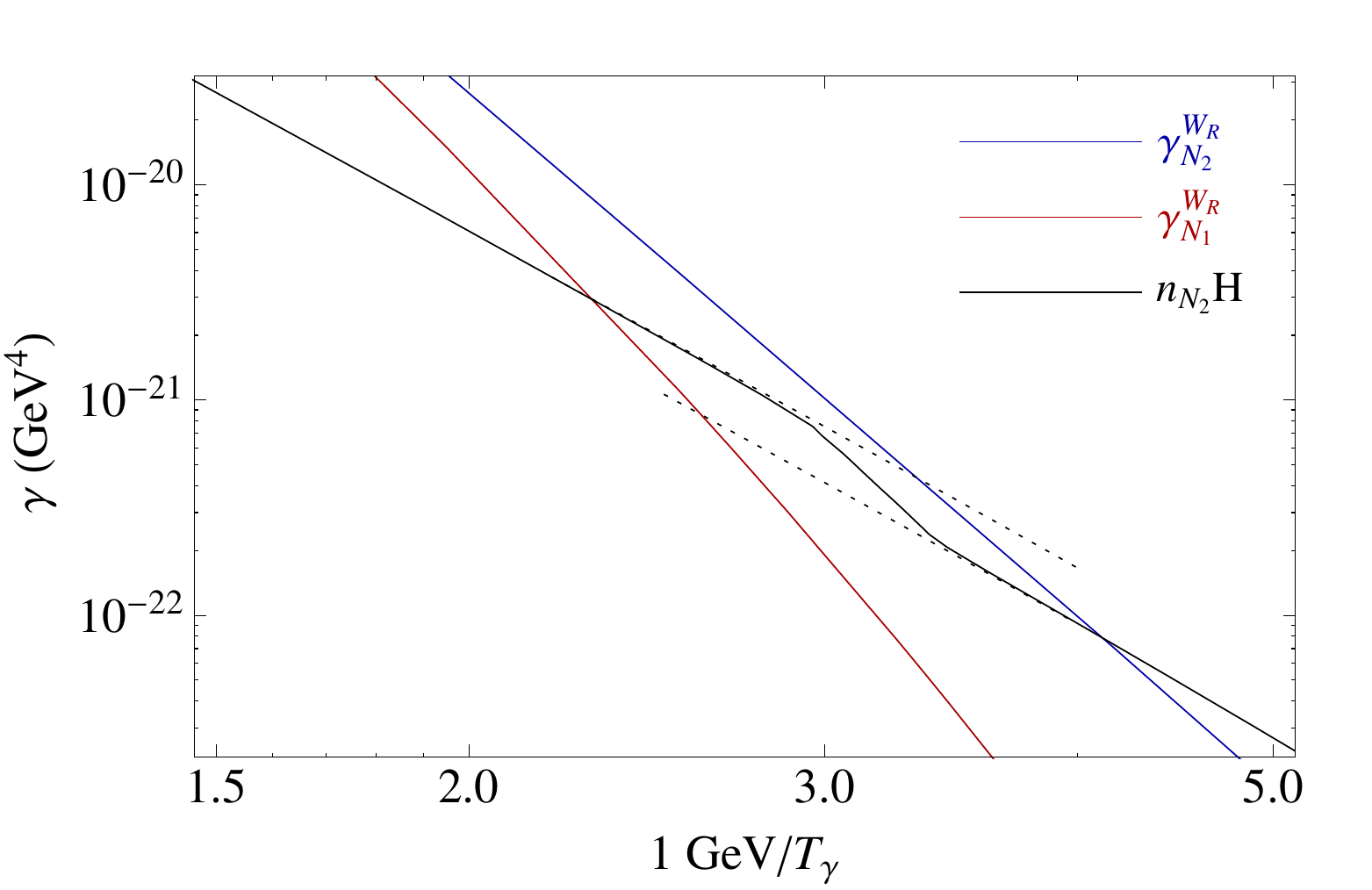}\hspace{-0.5cm}
\includegraphics[width=8.1cm]{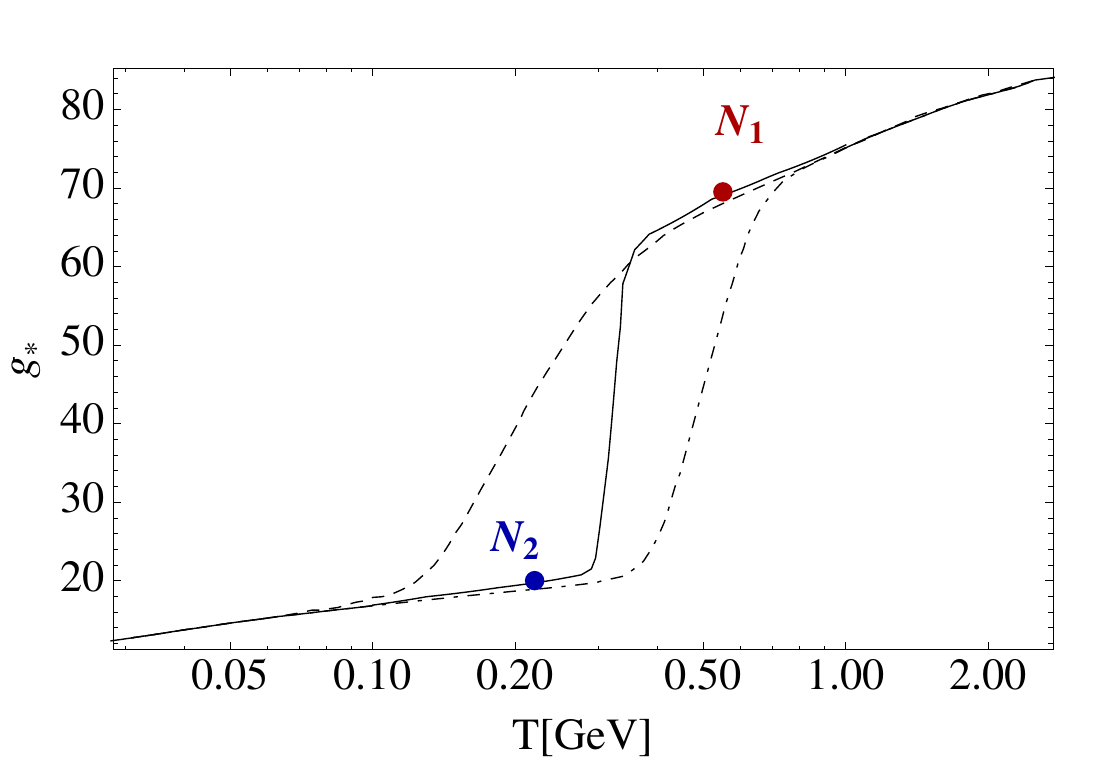}}
\caption{{\bf Left.} Thermally averaged reaction rates for the processes that dominate the decoupling of $N_1$ (red) and $N_2$ (blue), for parameters $M_{W_R}=5\,$TeV, $m_{N_2}=0.25$\,GeV. Also shown is the Hubble expansion rate multiplied by the thermal number density of $N_2$ (black curve). {\bf Right.}  A sharp change in the evolution of $g_{*S}$ around the QCD phase transition temperature. The dot-dashed (dashed) curve corresponds $T_{\rm QCD}=400 (150)\,$MeV with a second-order QCD phase transition (taken from~\cite{Srednicki:1988ce}), while the solid line is an interpolation in between with $T_{\rm QCD}=350\,$MeV and the transition close to first order. The red and blue points are the freeze-out temperatures of $N_1$ and $N_2$, respectively.}
\label{figGammaH}
\vspace{-1.ex}
\end{figure}

\subsection{A Window for low scale LRSM \label{secwindow}}

For obvious reasons, we here focus on the attractive possibility of having $W_R$ as light as possible. From Eq.~\eqref{dilutedmrelic}, it is clear that the ratio of $g_*(T_{f2,3})/g_*(T_{f1})$ should be small enough in order to compensate for the smallness of $m_{N}$. This amounts to decoupling $N_1$ as early as possible, compared to $N$. In the LRSM, the main processes that keep any RH neutrino in thermal equilibrium are 
\begin{equation}
	N \ell^- \to \bar u d, \ \ \  N u \to \ell^+ d, \ \ \  N \bar d \to \ell^+ u,
\end{equation}
mediated by $W_R$, and
\begin{equation}
	N  N \to \ell^+\ell^-,\ u \bar u,\ d \bar d,
\end{equation}
mediated by both $W_R$ and $Z_{LR}$. The relevant cross-sections and their thermally averaged reaction rates are given in the Appendix~\ref{AppndxXs2}. We find that the single-$N$ annihilation rate dominates the pair annihilation at least by one order of magnitude. The main reason is the suppression of the quark couplings to $Z_{LR}$, as mentioned above. 

A crucial point to note is that, in the single-$N$ annihilation processes, when the temperature drops below the mass of the corresponding charged lepton, the reaction rate starts to be Boltzmann suppressed, by more than one order of magnitude for temperatures around and below the QCD phase transition.~\footnote{There is a subtlety in the treatment of relevant annihilations at temperature close to the QCD phase transition. The issue is whether one should work with quarks or light mesons. Admittedly, the quark picture we opted for brings in uncertainties that cannot be easily quantified. In practice, since we ask the RH neutrinos to freeze out before or immediately after the transition, we believe such uncertainty is small.} Decoupling $N_1$ earlier than $N_2$ therefore requires a particular flavor structure of the right-handed leptonic mixing matrix $V^{R}_{\ell i}$. Namely, the $N_1$ should couple predominantly to the heaviest charged lepton, $\tau$, i.e. $V^R_{\tau1} \simeq 1$ (cf. Eq.~\eqref{flavorspectra}). In such a case, when the temperature drops below the $\tau$ mass, $N_1$ is maintained in the equilibrium only through the neutral current interactions, governed by $Z_{LR}$. On the other hand, $N_{2,3}$ now couples to $\mu$ or $e$ and keeps annihilating through the charged current interactions, too, and thus decouples later than $N_1$. In practice, the difference between the two decoupling temperatures can be as large as a few hundred MeV.
\begin{figure}[t]
\centering
\includegraphics[width=9cm]{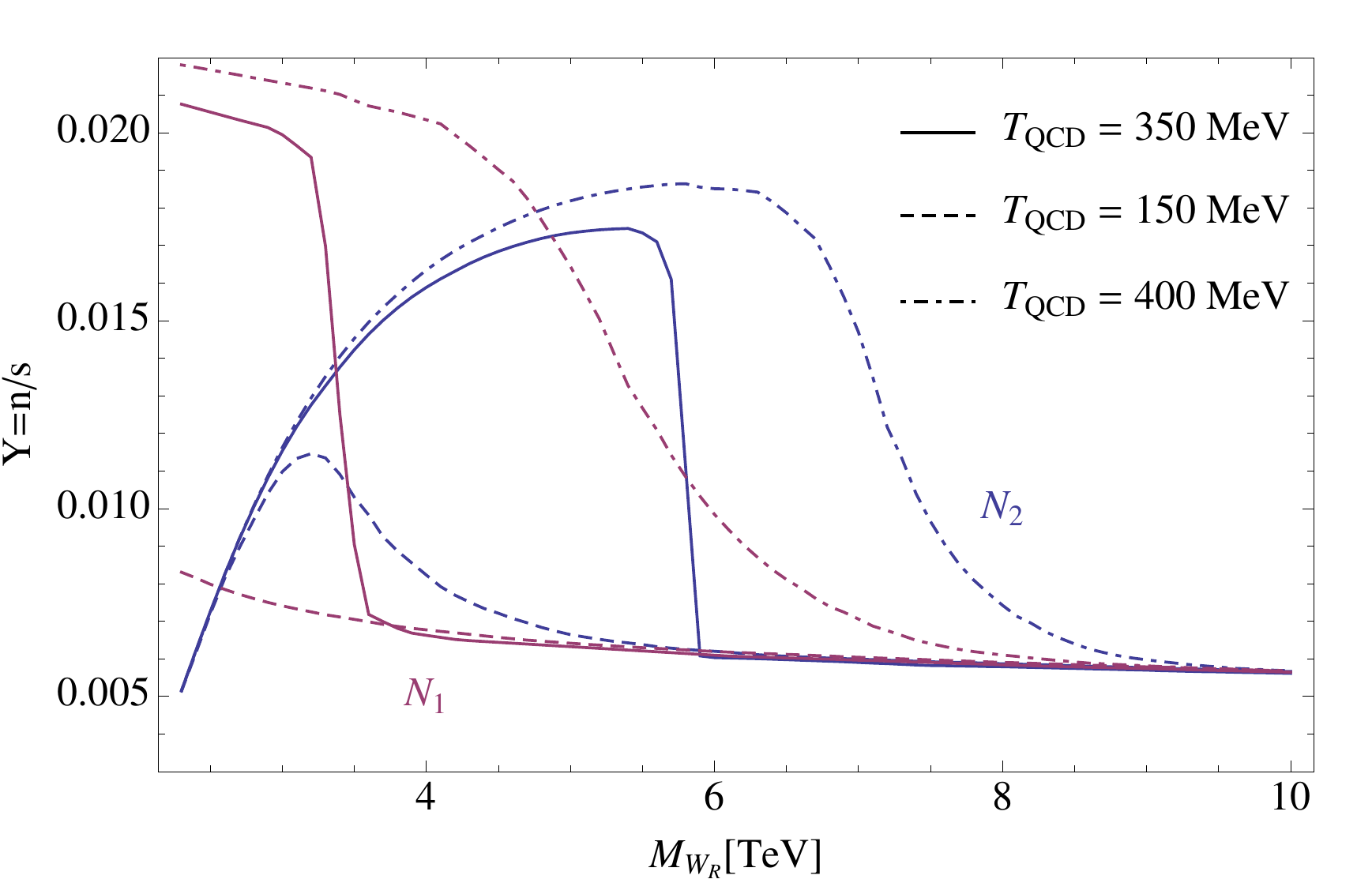}
\caption{The yield of $N_1(N_2)$ at freeze-out is shown in magenta (blue), depending on the mass of $M_{W_R}$. Different types of lines correspond to different QCD temperatures as denoted in the leggend. Notice that the nature of the QCD phase transition becomes irrelevant once $M_{W_R} \gtrsim 10\,$TeV, since the freeze-out temperature is above GeV. See also the right panel of Fig.~\ref{boltzsample} for more precise such dependence evaluated using Boltzmann equations .}
\label{figApprxY}
\end{figure}

Such a difference between the temperatures turns out to be important if the thermodynamical nature of the universe changes at the same time. For $W_R$ around a few TeV, the freeze-out temperatures of $N$ lie around a few 100 MeV. This is precisely in the vicinity of the QCD phase transition, where a sharp drop in $g_*$ occurs, as most hadron states are becoming non-relativistic (right panel of Fig.~\ref{figGammaH}). In case the total entropy is conserved during the transition, the temperature of the thermal plasma gets ``reheated" due to the change of $g_*$. 

For the sake of illustration, we show first what happens in a simple example with a single diluter, which we choose to be $N_2$ due to its larger mass and a bigger impact on the dilution factor. If the left-right scale is chosen, such that $N_1$ freezes-out before and $N_2$ after the phase transition (see Fig.~\ref{figGammaH}), $N_2$ will feel this reheating which will enhance its yield. The relative number density between two relativistic RH neutrinos is
\begin{equation}\label{YNratio}
  \frac{Y_{N_2}}{Y_{N_1}} \sim \frac{g_{*}(T_{f1})}{g_{*}(T_{f2})} \ ,
\end{equation}
and can be as large as $3-4$ for $M_{W_R}\lesssim10 \text{ TeV}$, as shown in Fig.~\ref{figApprxY}. This will eventually result in more entropy release during the later decay of $N_2$, and it is manifest in the reduction of the ratio ${g_{*}(T_{f2})}/{g_{*}(T_{f1})}$ in Eq.~\eqref{dilutedmrelic}. Although in this simplified picture the enhancement factor is not yet large enough to completely compensate for the smallness of $m_{N_2}$ as a diluter mass in Eq.~\eqref{dilutedmrelic}, we show below that when both $N_2$ and $N_3$ are introduced into the game, the dilution factor $S$ can be large enough for the correct DM relic abundance. Depending on the temperature of the QCD phase transition, there is a window of $M_{W_R}$ for this effect to be significant. This relatively low scale window will be quantified in the next section to lie around 5 TeV or so; for the time being the essential point is just its existence. It is only a window, since for even smaller $M_{W_R}$, below $\sim$\,3 TeV, the freeze out temperature is below $N_2$ mass, and the dilution factor gets Boltzmann suppressed.

\subsection{Diluters and dilutees: summary \label{secPhaseDiagram}}

\begin{figure}[t]
  \hspace{-0.2cm}
  \centerline{
  \includegraphics[width=9.5cm]{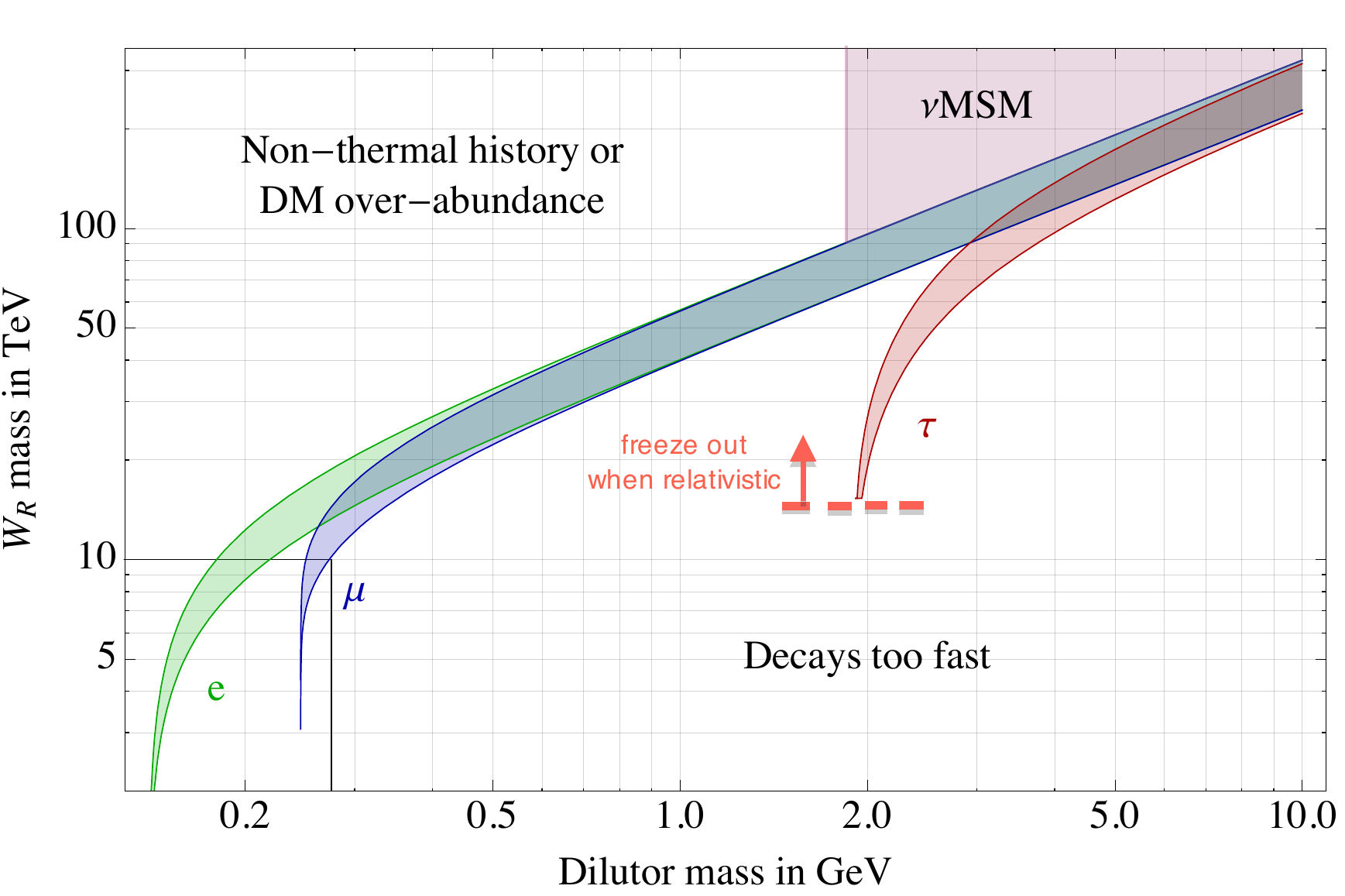}
  \raisebox{13pt}{\includegraphics[width=5.05cm]{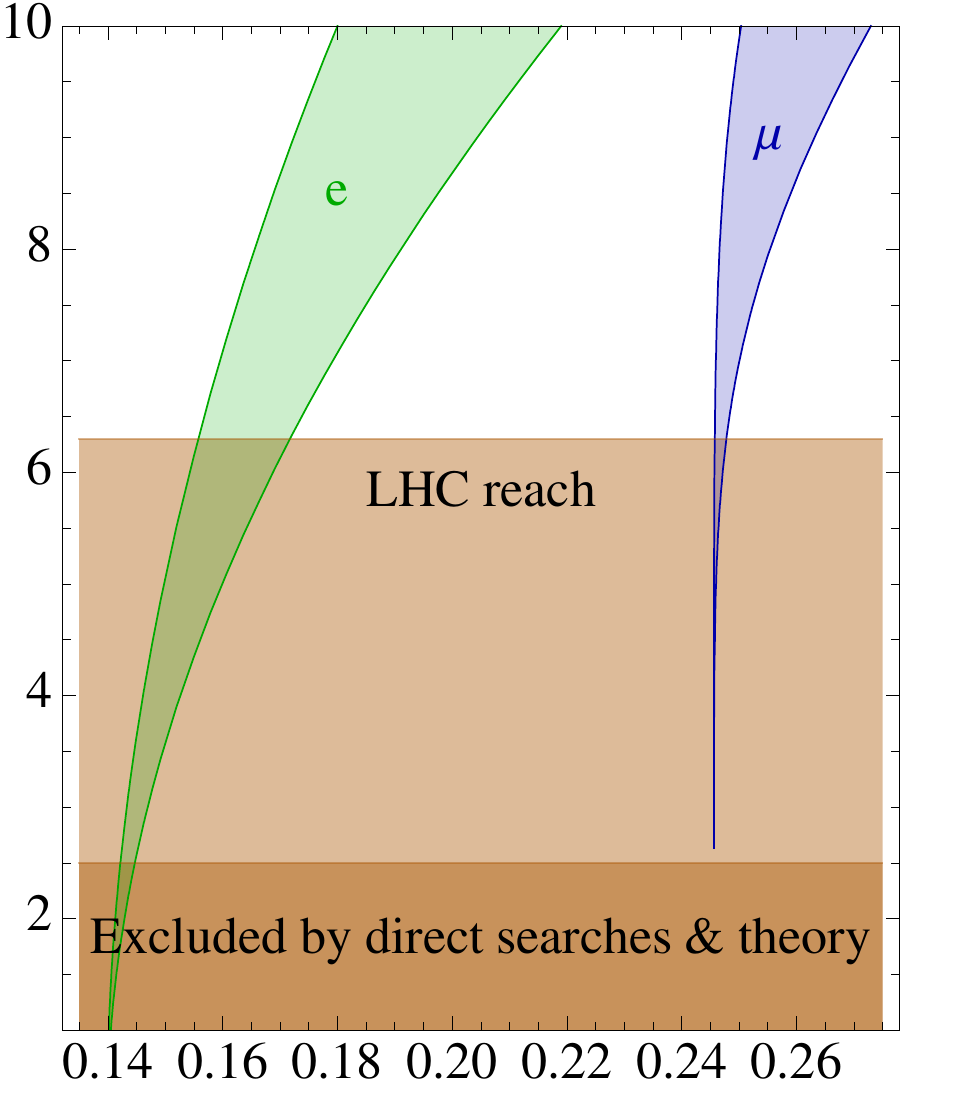}}
  }
  \caption{Parameter space relevant for the warm dark matter in the minimal LR model. {\bf Left.} The shaded regions (green, blue and red) labeled $e, \mu$ and $\tau$ depict the regions, where the lifetime of the diluting $N$ lies between 0.5 and 2 seconds, with $N$ coupling predominantly to a single flavor. The spikes due to the phase space suppression in the decay of $N$ allow for lower values of $M_{W_R}$. {\bf Right.} A zoom to the region $M_{W_R}\lesssim10\,$TeV of our primary interest. Also shown are the theoretical lower limit on $M_{W_R}$ from kaon mixing (which coincides with the lower limit set by the current LHC direct search of $W' \to e/\mu$ + missing energy), as well as the 14 TeV LHC reach.
  \label{figPhaseSpace}}
\end{figure}

This is an appropriate moment to summarize our findings in a single spot and take a look at the parameter space of the LRSM in view of the relic density of the RH neutrino $N_1$, which we show in Fig.~\ref{figPhaseSpace}. The shaded belt regions show the possible parameter space where significant entropy production has a chance to take place. They correspond to the lifetime of the diluters within the 0.5\,--\,2 second range. If they are much heavier than GeV, they decay too fast, unless $W_R$ is heavy ($\gtrsim20\,$TeV). In this regime, the parameter space has a simple scaling law $m_N \propto M_{W_R}^{4/5}$. Notice that only when $m_N \approx m_\pi + m_\ell$, lower values of $M_{W_R}$ are allowed, producing the spikes seen in Fig.~\ref{figPhaseSpace} in accord with the flavor diagonal $\mathbf{V^R_\ell}$ in Eq.~\eqref{flavorspectra}.

Furthermore, the above regions terminate at values of $M_{W_R}$ for which the diluters start to feel the non-relativistic suppression during its freeze out. In the case of $\ell=\tau$, this gives a lower bound $M_{W_R}\gtrsim 16\,$TeV~\cite{Bezrukov:2009th}. Once again, this is the reason why we focus here on lighter diluters which do not couple to $\tau$ in order to have LR symmetry near the TeV scale. As we showed above, and as will be discussed in detail in the next technical session, {\it the main message here is the existence of a window
of the LR scale, potentially accessible to the LHC}, besides the expected high scale scenario discussed in the past.

Within the region below the belts, the decays of diluters are usually too fast and one ends up having too much DM in the universe. For the region above, however, the decays are not necessarily slow, because there is still a possibility to decay via the Dirac Yukawa couplings, which we have neglected so far. These couplings induce the mixing between the RH and SM neutrinos, and new decay channels $N \to 3\nu$ or $\nu e^+e^-$ mediated by the SM gauge bosons open up. If these decay channels dominate over the $SU(2)_R$ gauge interactions, the production and dilution of $N_1$ crosses over smoothly to the $\nu$MSM~\cite{Asaka:2005an, Asaka:2006ek} case in the phase diagram.

However, there is a subtle difference from the usual $\nu$MSM picture due to the presence of new gauge interactions, which now enter the game. Even though they do not play any role in the decay of the diluter, they may still thermalize and over-produce the DM candidate $N_1$, if the universe started out at a high enough temperature. In this case, a late-time dilution is still necessary, which calls for $m_N \gtrsim$\,1\,--\,2\,GeV~\cite{Asaka:2006ek}. This corresponds to the magenta shaded region in Fig.~\ref{figPhaseSpace}. Another way out would be to consider the reheating temperature of the universe after inflation to be sufficiently lower than the typical freeze-out temperature on the order 100 MeV - GeV. In this case, $N_1$ has to be produced in a non-thermal way~\cite{Dodelson:1993je, Shi:1998km, Abazajian:2001nj}.

%
%
\section{The Boltzmann Approach \label{secN2N3Delta}}

As promised, we now come to the section of the technical aspects of our computations. Here, we implement the picture described in the previous section and numerically solve for the dilution of DM relic abundance. This quantifies the window of low scale LR symmetry consistent with a warm DM candidate. If not yet done, this is the right moment for our reader to take a (short) coffee break.

In order to set the stage, let us recapitulate the main points which follow from the qualitative study of the previous section.
\begin{enumerate} 
\item The lightest neutrino $N_1$, the DM candidate, weighs around a keV, and therefore is always relativistic throughout the thermal history of interest here. It is coupled to $\tau$-lepton only ($V^R_{\tau 1}=1$), which suppresses the charged current interactions when the temperature drops below the $\tau$ mass, leaving room only for neutral currents mediated by $Z_{LR}$.
\item Such a flavor structure guarantees that a diluter $N$ does not decay to $N_1$ if its mass is below $m_\tau$.
\item There is a profound difference between the decoupling of the diluters $N$ and the light neutrinos (and $N_1$). A diluter's freeze-out temperature is determined by the single $N$ scattering with quarks/pions through charge-current interactions.
\item The late decay of a diluter happens at the temperature around MeV, which is well below their freeze-out temperature and therefore we are going to treat these processes in two separate stages.
\item In order to optimize the entropy production, the diluters with lifetimes as long as a second should be as heavy as possible. In turn, this implies $V^R_{\mu 2} \sim V^R_{e 3} \approx 1$ and $m_{N_2} \approx m_{\pi}+m_\mu, m_{N_3} \approx m_\pi+m_e$. In this case, the pionic decay dominates and the $M_{W_R}$ dependence is shown in Fig.~\ref{figPhaseSpace}.
\end{enumerate}

\subsection{Freeze out}

We now turn to the study of the Boltzmann equations governing the freeze-out of RH neutrinos. In order to keep track of the expansion of the universe, we define an arbitrary temperature $T$, which simply scales as $1/R$. One can think of $T$ as the temperature of some fictitious relativistic species which freezes-out at some initial temperature $T_i$. Notice that the photon temperature is no longer a convenient choice as the number of relativistic degrees of freedom, i.e. $g_*$ changes dramatically with time, especially during the QCD phase transition. In any case, one can always solve for the photon temperature $T_\gamma$ in terms of $T$, using entropy conservation
\begin{equation}
	g_*(T_i) \, T^3 = g_*(T_\gamma) \, T_\gamma^3 \ .
\end{equation}
In practice, we choose the initial temperature to be $T_i=10\,$GeV and find $T<T_\gamma \lesssim 2 \,T$ throughout the freeze out process of $N$'s.

The set of Boltzmann equations describing the freeze out of $N_i$ are
\begin{equation} \label{boltzmann3N}
s H z \frac{d Y_{N_i}}{d z} = - \left[ \frac{Y_{N_i}}{Y_{N_i}^{eq}} -1 \right]  \gamma^{W_R}_{N_i} - \left[ \left(\frac{Y_{N_i}}{Y_{N_i}^{eq}}\right)^2  -1 \right]  \left(\gamma^{Z_{LR}}_{N_iN_i} + \gamma^{W_R, Z_{LR}}_{N_iN_i} \right), i=1,2,3 \ , 
\end{equation}
where~\footnote{We choose $m_{N_2}$ as the normalization point, since $N_2$, being heavier than $N_3$, is the dominant source of dilution.} $z\equiv m_{N_2}/T, z_\gamma \equiv m_{N_2}/T_\gamma$ and the entropy density and the Hubble parameter are defined as usual
\begin{equation}
	s = \frac{2 \pi^2}{45} g_{*S} T_\gamma^3, \quad H = 1.66 \sqrt{g_{*}} \frac{T_\gamma^2}{M_{\text{p}}} \ .
\end{equation}
The single- and pair-annihilation reaction rates $\gamma_{N_i}^{W_R}, \gamma_{N_iN_i}^{W_R, Z_{LR}}$ and $\gamma_{N_iN_i}^{Z_{LR}}$ are given in the Appendix~\ref{AppndxXs2}. As discussed in the section above, the single-$N$ processes mediated by $W_R$ dominate by far over the pair annihilation processes. The latter interactions only affect the result by less than 5\%. For LR symmetry scale close to TeV, the charged-current interaction rate of $N_1$ is relatively smaller compared to that of $N_{2,3}$, due to the Boltzmann suppression in the $\tau$-lepton number density. 

\begin{figure}[t]
\centerline{\includegraphics[width=8.3cm]{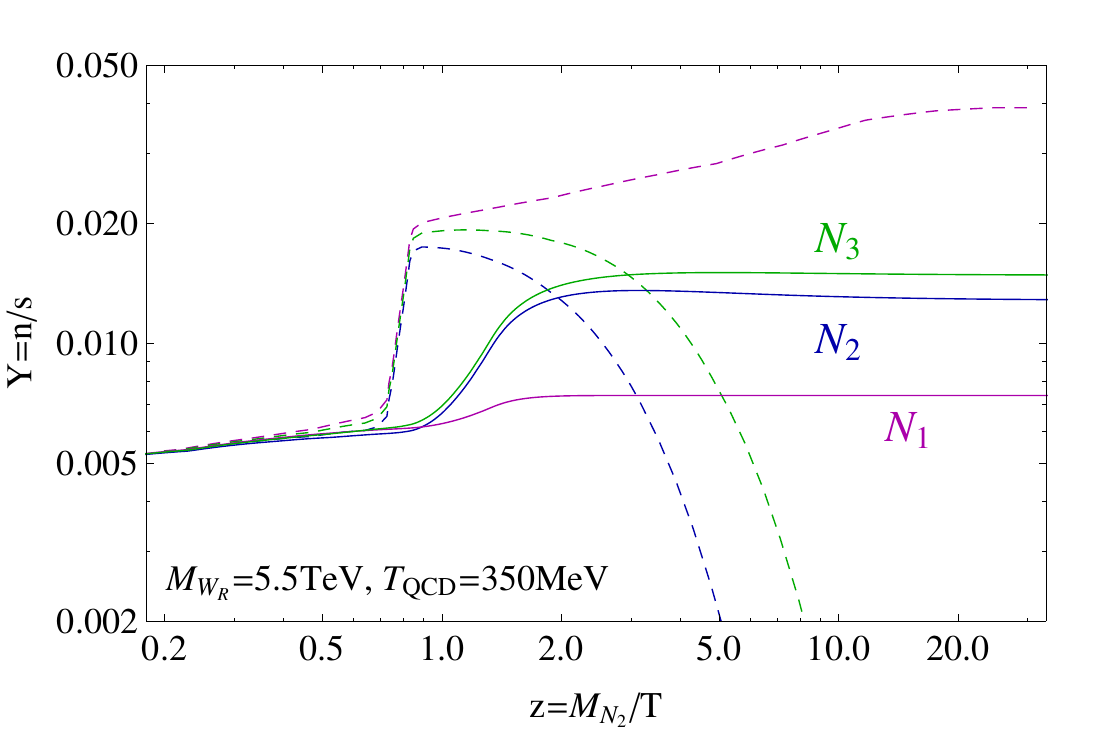}\hspace{-0.5cm}
\includegraphics[width=8.5cm]{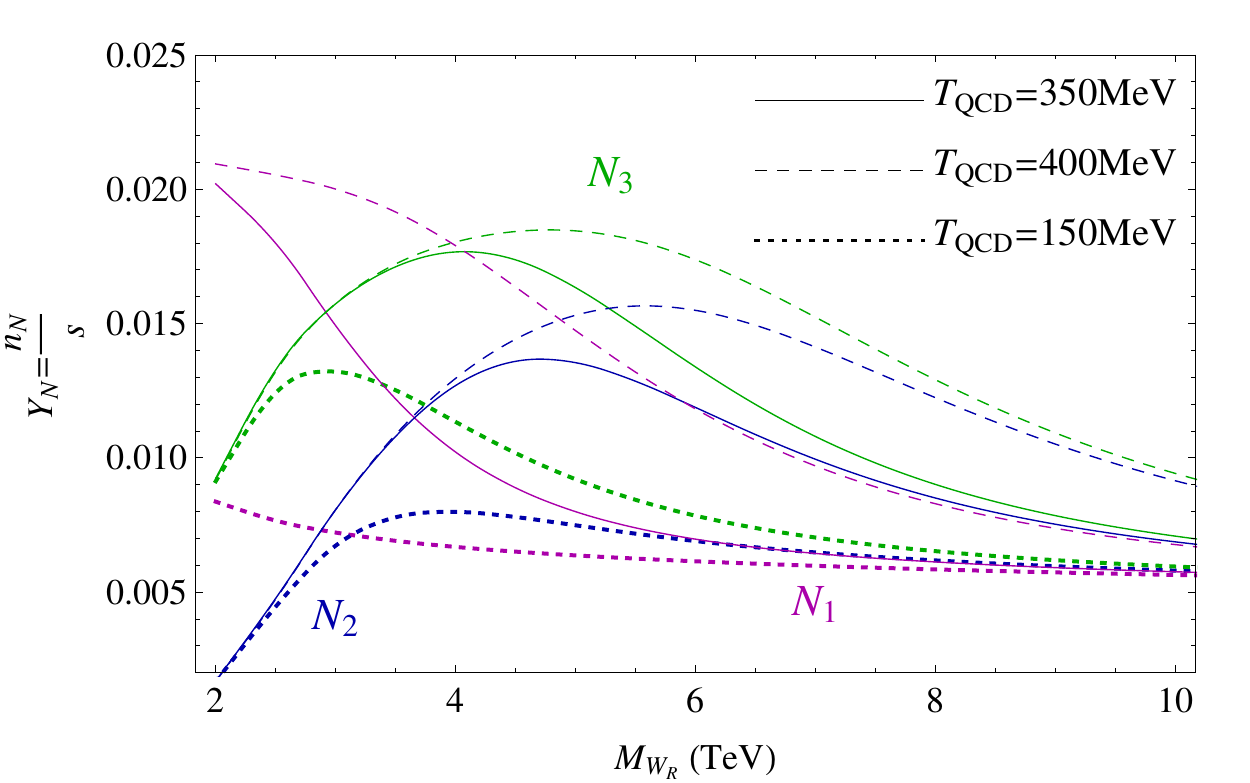}}
\caption{Solid lines representing the yields of $N_1$ (magenta), $N_2$ (blue) and $N_3$ (green) during the freeze out process, as solved using the Boltzmann equations in Eq.~(\ref{boltzmann3N}). In both panels, the masses of $N_{2,3}$ correspond to the lifetime of $1.5 \text{ sec}$. {\bf Left.} Yields of $N_i$ as a function of temperature during the freeze-out process for a fixed value of $M_{W_R} = 5.5 \text{ TeV}$. The dashed curves corresponds to the thermal equilibrium yields of $N_i$, while the kinks reflect the reheating effect of QCD phase transition.
{\bf Right.} The reader may wish to take a deep breath before staring at the this panel. Final yields for different LR scales as a function of $M_{W_R}$ computed using different values of $T_{\text{QCD}}$ as explained in the legend (cf. with the naive estimate in Fig.~\ref{figApprxY}).  }
\label{boltzsample}
\end{figure}

We have numerically solved the above Boltzmann equations to calculate the yields of the three RH neutrinos. We start evolving the equations from a sufficiently high temperature, where all the states are in thermal equilibrium. A sample solution of the yields depending on $z$ (temperature) during freeze-out is shown in the left panel of Fig.~\ref{boltzsample}, where we take $M_{W_R} = 5.5 \text{ TeV}$, $T_{\text{QCD}}=350\,$MeV and we obtain masses of $N_{2,3}$ by fixing their lifetime equal to $1.5 \text{ sec}$. In this case, $N_1$ freezes out just before the QCD phase transition and its temperature barely receives the heating from $g_*$ change. In contrast, $N_2$ and $N_3$ can freeze out after the transition and before they become non-relativistic, thus their number densities get enhanced. This realizes the large relative ratio discussed in Eq.~\eqref{YNratio}.

We proceed to compute the dependence of the final yields for a varying LR scale and show $Y_{N_i}$ as a function of $M_{W_R}$ in the right panel of Fig.~\ref{boltzsample}. In order to obtain the final yields, we take the numerical solution at $z \approx 30$, where the freeze-out is basically finished, and where the time is still early enough so that $N_{2,3}$ have not yet started to decay. The final yields obtained here will serve as initial conditions for the Boltzmann equations describing the late decay of $N_{2,3}$ in the following section.

The realistic numerical results of this section are to be compared with the estimate in Fig.~\ref{figApprxY}. One can see that the largest ratios of $Y_{N_{2,3}}/Y_{N_1}$ can be achieved for $M_{W_R}$ between $4-6 \text{ TeV}$. During the freeze-out process, even though the interaction rate has dropped below the Hubble rate, its effect on the evolution does not disappear immediately. This is why the final ratio tends to be smeared compared to the rough estimate in Fig.~\ref{figApprxY}.

\subsection{Late decay}

The calculation of entropy production due to the late decay of diluters $N_{2,3}$ is based on the assumption that the microscopic interactions of the remaining SM particles in the thermal plasma are still fast enough. This guarantees that the relaxation time for the decay products to equilibrate with the plasma is much shorter than the time scale of the universe expansion. Therefore, the heat release $dQ$ from the decay is completely transferred into the energy of radiation. The final entropy release $\Delta S = \int dQ/T$~\cite{Scherrer:1984fd} can be obtained by studying the Boltzmann equations for the energy density evolution~\cite{Nagano:1998aa}. In our case, the typical range of temperatures for the matter domination and decay to happen is between $0.5 -10 \text{ MeV}$. The set of relevant Boltzmann equations for the energy density of matter and radiation are:
\begin{align}
  \frac{d \rho_R}{dt} + 4 H \rho_R &= \Gamma_2 \rho_2 + \Gamma_3 \rho_3 \ ,  
  \\
  \frac{d \rho_{N_1}}{dt} + 4 H \rho_{N_1} &= 0 \ , 
  \\
  \frac{d \rho_{N_2}}{dt} + 3 H \rho_{N_2} &= - \Gamma_2 \rho_2 \ ,
  \\
  \frac{d \rho_{N_3}}{dt} + 3 H \rho_{N_3} &= - \Gamma_3 \rho_3 \ .
\end{align}
The initial conditions are obtained by matching this late decay regime to the freeze-out regime in the previous section, at time $t_{\rm m}$
\begin{align}
\rho_{\gamma}(t_{\rm m})& = \frac{\pi^2}{30} \, g_{*}(T_{\gamma, \rm m}) \, T_{\gamma, \rm m}^4 \ , 
\\ 
\rho_{N_1}(t_{\rm m}) &= \frac{7}{4} \, \frac{\pi^2}{30}\, T_{N_1, \rm m}^4 \ , 
\\
\rho_{N_2}(t_{\rm m}) &= m_{N_2} \, Y_{N_2}(z_{\rm m}) \, \frac{2 \pi^2}{45}\,g_{*}(T_{\gamma, \rm m}) \,T_{\gamma, \rm m}^3\ , 
\\ 
\rho_{N_3}(t_{\rm m}) &= m_{N_3} \, Y_{N_3}(z_{\rm m}) \, \frac{2 \pi^2}{45}\,g_{*}(T_{\gamma, \rm m}) \,T_{\gamma, \rm m}^3 \ .
\end{align}

We choose a matching point with $z_{\rm m}=30$, or photon temperature around 15\,MeV, where the temporary matter domination of diluters has not yet begun. The corresponding Hubble time can be calculated using $t_{\rm m}=1/2H(T_{\gamma, \rm m})$. We evolve the above equations up to a time $t_{\rm fin}$ when the entropy production has finished, typically much larger than the lifetime of a diluter.

The reheating temperature $T_r$ after the decay is defined as the temperature when radiation again starts to dominate the universe, which happens around $t = \tau_N$. It can be extracted from the radiation energy density at that time
\begin{equation}
  \rho_{R}(\tau_N) = \frac{\pi^2}{30} g_*(T_r) T_r^4 \ .
\end{equation}

The dilution factor $\mathcal{S}$ is defined as the increase of total entropy $S$ due to the late decay, as was 
discussed at length in Section~\ref{secTwoNs}. It can be derived based on the fact that radiation-like SM particles absorb all the heat from $N_{2,3}$ decay, therefore the products of the diluters dominate by far the entropy density. On the other hand, our DM $N_1$ is by now completely decoupled from the plasma and since it is not a product of the heavier neutrinos, it simply dilutes as $\rho_{N_1} \propto R^{-4}$. One then gets an improved dilution factor, which was estimated before in Section~\ref{secTwoNs} in the sudden decay approximation (see Eq.~\eqref{eqDilutionS})
\begin{equation}
 \mathcal{S}_{\rm improved} = \frac{S(t_{\rm f})}{S(t_{\rm m})} = \frac{s(t_{\rm f})}{s(t_{\rm m})} \frac{V(t_{\rm f})}{V(t_{\rm m})} = \left[ \frac{\rho_R(t_{\rm f})}{\rho_R(t_{\rm m})} \right]^{3/4} \left[ \frac{\rho_{N_1}(t_{\rm f})}{\rho_{N_1}(t_{\rm m})} \right]^{3/4} \ .
\end{equation}
This dilution factor will be used to rescale the relic abundance of $N_1$ in Eq.~\eqref{dmrelic}, where $Y_{N_1}$ is also calculated numerically from the previous subsection.
 
\begin{figure}[t]
\centering
\includegraphics[width=10cm]{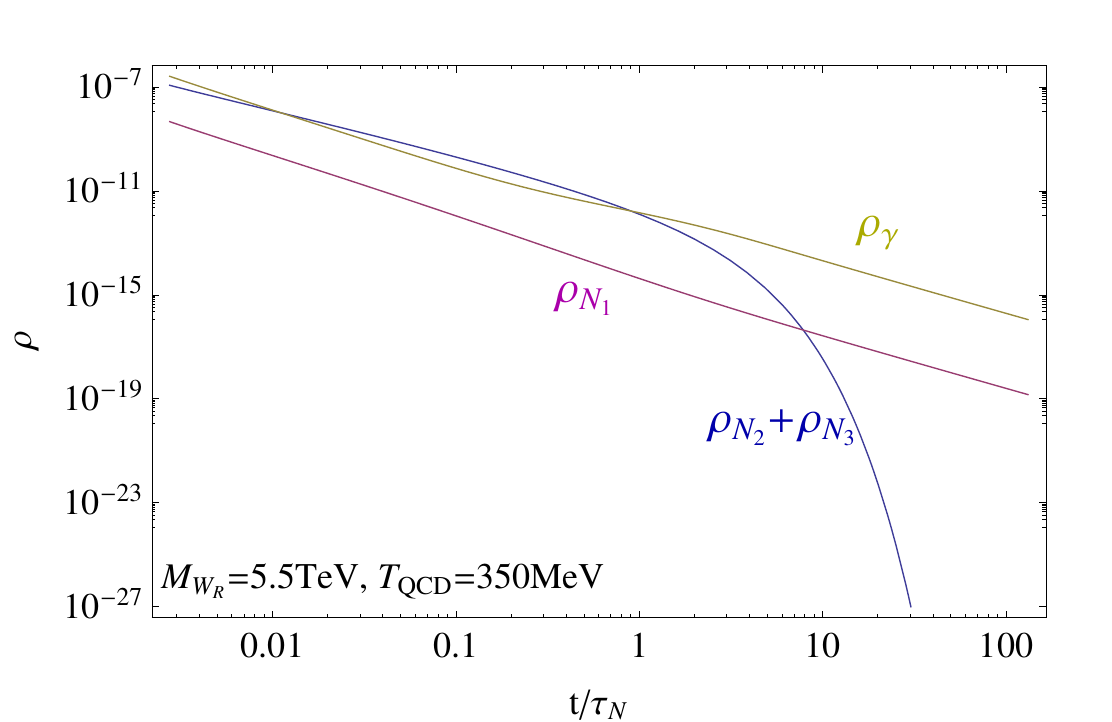}
\caption{Time evolution of energy densities for different species in the universe. We have used the same set of parameters as the left panel of Fig.~\ref{boltzsample}. 
The universe temporarily enters a matter-dominated phase before $N_{2,3}$ decay. }\label{rhoNs}
\end{figure}

A sample solution with model parameters $M_{W_R}=5.5\,$TeV, $T_{\text{QCD}}=350 \text{ MeV}$ and $\tau_{N}=1.5$\,sec is shown in Fig.~\ref{rhoNs}. The initial conditions are obtained by matching to the sample values obtained from the freeze-out. We find the energy density of $N_{2,3}$ can temporarily dominate over that of radiation before their decay, and the universe temporarily enters matter dominated expansion. The matter domination factor $(\rho_{N_2}+\rho_{N_3})/\rho_R$ can be as large as 3. In this case, the reheating temperature is $T_r\approx 0.7\,$MeV and  could be consistent with helium abundance in standard BBN~\cite{Kawasaki:2000en}. We comment on this in more detail, as well as on the implications for the CMB in the following section.

\begin{figure}[t] \centerline{
\includegraphics[width=8.5cm]{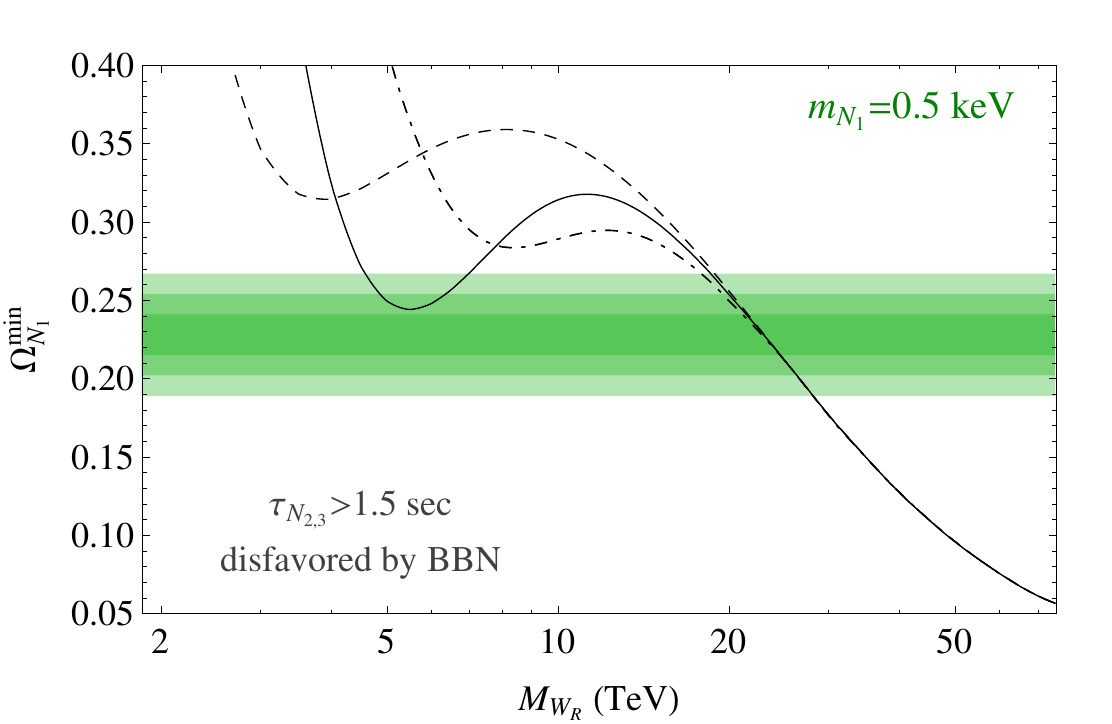}
\includegraphics[width=8.5cm]{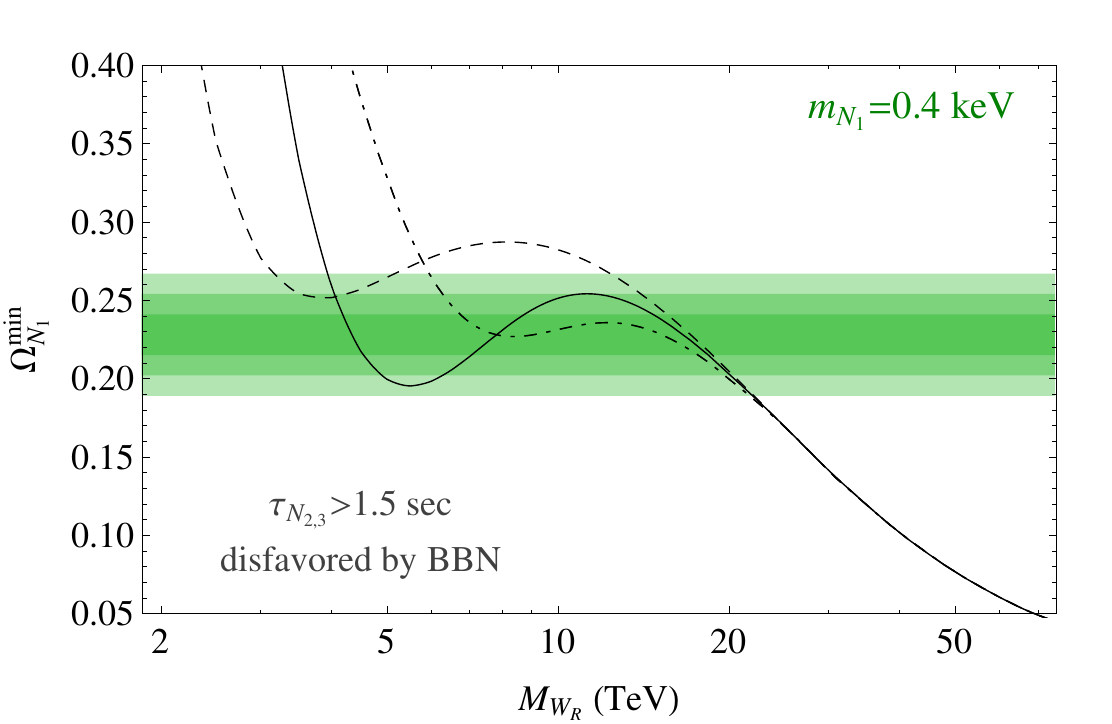}}
\caption{The minimum DM relic density that can be accommodated in the LR theory is plotted for two fixed DM masses as a function of $M_{W_R}$. Solid, dashed and dot-dashed lines correspond to $T_{\text{QCD}} = 350 \text{ MeV}$ (nearly first order) and $T_{\text{QCD}} = 150, 400\,\text{MeV}$ (second order), respectively and a fixed lifetime $\tau_{N_{2,3}} = 1.5 \text{ sec}$. The green bands from dark to light correspond to DM relic abundance at 1, 2, 3\,$\sigma$ confidence level from WMAP fit. Keep in mind that the DM relic density is linearly proportional to $m_{N_1}$.}\label{figRelicDensity}
\end{figure}

In order to quantify the low-scale $M_{W_R}$ window in which the relic density of $N_1$ agrees with that of dark matter, we do a complete numerical study and compute relic density $\hat \Omega_{N_1}$ after the maximal dilution, with $\tau_{N}=1.5\,$sec or $T_r\approx 0.7\,$MeV. In Fig.~\ref{figRelicDensity}, $\hat \Omega_{N_1}$ is plotted in together with the WMAP favored value $\Omega_{\rm DM}$ (Eq.~\eqref{pdgrelic}) shown in green bands.
For a varying lifetime, the reheating temperature scales as $1/\sqrt{\tau_{N}}$, while the dilution factor $\mathcal{S}$ scales as $\sqrt{\tau_N}$. We find that the desired window exists for $m_{N_1}=0.5\,$keV. In the case of nearly first-order QCD phase transition happening at 350\,MeV, the light $W_R$ window lies between $4-8 \text{ TeV}$, while for a second-order transition at around 400\,MeV, the window shrinks and shifts to around $8 - 9 \text{ TeV}$. For lower QCD transition temperatures, the window moves to lower $M_{W_R}$, but calls for an even lighter DM mass $m_{N_1} \sim 0.4 \text{ keV}$. For such a value, the window is there, regardless of the nature of the QCD phase transition.

\subsection{Dark Matter and LR scale: summary}

Here we summarize our quantitative results after numerically scanning over the parameter space. In Fig.~\ref{figRelicDensity} we plot the {\it minimal} relic density of the DM candidate $N_1$ as a function of the LR symmetry scale $M_{W_R}$ for two different values of $m_{N_1}$, together with the allowed values coming from observations. Let us explain the qualitative aspects of the plot on the left panel of the figure, for the more conservative value $m_{N_1} = 0.5 \text{ keV}$.

First of all, the reader should recall that it was crucial to decouple our DM $N_1$ as early as possible in order to separate its freeze-out temperature from the one of the diluters, profiting from the QCD phase transition. This is why $N_1$ had to be coupled to $\tau$ in order to have predominantly weaker neutral interactions at that point in thermal history. In order for this to work, $W_R$ can not be too light;  otherwise the $T_f$'s of all $N$'s are equal and below $T_{\text{QCD}}$ and $N_1$ remains over-abundant.


Now for the window around $M_{W_R} \approx 5 \text{ TeV}$ obtained by successfully separating $T_f$ of the diluters and the dilutee. As seen in the figure, this works only for the nearly first-order QCD phase transition happening at around 350 MeV. If the DM mass is allowed to be this low, it is remarkable that the $W_R$ mass window lies at the heart of the 14 TeV LHC reach~\cite{Ferrari:2000sp, Gninenko}. 

Further increase in the $W_R$ mass results in the merging of these two different temperatures, now above $T_{\text{QCD}}$. As we go along, taking bigger values of $W_R$ mass, the relic DM abundance keeps falling down due to the fact that the mass of the diluters, directly controlling the dilution, gets increased, see Fig.~\ref{figPhaseSpace}.

The DM relic density simply scales linearly with $m_{N_1}$, therefore if one were allowed to go to even smaller values of $m_{N_1}$, say 0.4 keV as in the right panel, the correct DM abundance could be achieved for basically any value of $T_{\text{QCD}}$. In such case the $W_R$ mass window either expands or the upper boundary completely disappears.

For the sake of completeness, we wish to remind the reader that there was also the other possibility of $N_1$ being coupled to light leptons, but this would force $W_R$ to be heavy, definitely above the LHC reach and therefore not of our interest here.

%
%
\section{Further Constraints and the Uncertainties \label{secConstraints}}

We now turn to astrophysical, cosmological and low energy constraints on the DM scenario discussed above. They can be classified into constraints on the DM mass $m_{N_1}$, lifetime of diluters $\tau_{N}$ and the left-right symmetry scale $M_{W_R}$. These bounds are summarized in Table~\ref{WDMBound}.

\subsection{Dwarf spheroidal galaxies and Lyman-$\alpha$ forest}

Cosmological observations put a lower limit on the mass of the DM candidate. The most reliable and conservative bound comes from the study of dwarf spheroidal galaxies, whose content is commonly believed to be dominated by dark matter. If fermionic DM inside such astrophysical objects can be regarded as a degenerate Fermi gas, the requirement that the DM velocity on the Fermi surface must be less than the escape velocity, gives a lower bound on its mass $m_{\rm DM} > 0.468^{+0.137}_{-0.082}$\,\,keV~\cite{Boyarsky:2008ju}. A more sophisticated analysis which compares the maximum phase space density~\cite{Tremaine:1979we, Boyarsky:2008ju, Gorbunov:2008ka} with observations gives a slightly stronger bound, $m_{\rm DM}>0.557^{+0.163}_{-0.097}$\,\,keV. Our case with $m_{N_1}\approx 0.5 \text{ keV}$ is consistent with these lower limits.

On the other hand, a lower limit on the DM mass can also be inferred from studying the absorption lines in the Lyman-$\alpha$ forest, which mainly constrains the maximal free-streaming length of warm DM. Most N-body simulations are carried out for structures formed at redshifts approaching the non-linear growth regime, and are subjected to large uncertainties. The most recent constraints are made for warm RH neutrinos produced via non-resonant production, with $T_{N_1}\sim T_{\nu}$, with a range between 8\,--\,14\,\text{keV}~\cite{Seljak:2006qw, Boyarsky:2008xj}. In our scenario, $N_1$ freezes out earlier than SM neutrinos and gets further diluted. In such case, its temperature, free-streaming length and the mass lower bound should be reduced by a factor of $\left(g_*(T_{f, \nu})/(g_*(T_{f1}) \mathcal{S}) \right)^{1/3}$. Therefore, the lower bound for our case is $m_{\rm DM}\gtrsim \mathcal{O}(1)$\,keV. We also notice that a recent analysis~\cite{deVega:2009ku, Menci:2012kk, Destri:2012yn} using a thermal relic warm DM with mass 0.75\,keV finds consistency with WMAP and Lyman-$\alpha$ observations. In view of the uncertainties mentioned above, we stick to our conservative lower limit about 0.5\,keV.

\begin{table}[t]
\centering
\begin{tabular}{|c|c|c|c|}
        \hline
	Constraints & $m_{N_1}$ & $\tau_{N}$ & $M_{W_R}$ \\ 
	\hline \hline
	Dwarf Galaxy & $\gtrsim 0.4-0.5\,$keV & --- & --- 
	\\ \hline
	Lyman-$\alpha$ & $\gtrsim 0.5$\,--\,$1\,$keV & --- & --- 
	\\ \hline
	BBN \& CMB & --- & $\lesssim 1.5\,$sec & --- 
	\\ \hline
	$0\nu2\beta$ & --- & --- & $\gtrsim6-8\,$TeV 
	\\ \hline
	LHC-14 reach & $0-M_{W_R}$ & --- & $\lesssim 6.3\,$TeV 
	\\ \hline
	A sample point & 0.5\,keV & 1.5\,sec & $4-7$\,TeV \\
	\hline
\end{tabular}
\caption{Various constraints on the masses and liftetimes of relevant states within the LRSM, coming from astrophysical, cosmological and terrestrial experiments, together with a sample point in the DM scenario. \label{WDMBound}}
\end{table}

\subsection{CMB and BBN: neutrino thermalization}

Another class of cosmological constraints is related to the reheating temperature after the late decay of $N_{2,3}$. As we learned in this study, a large enough dilution implies a longer lifetime of $N$ and therefore a lower reheating temperature. If this temperature drops below the SM neutrino decoupling temperature, which is around $1-2\,\text{MeV}$, the thermalization of neutrinos becomes inefficient. In turn, this has a strong impact on both, the CMB power spectrum and the production of light elements during BBN. Naively, one could conclude that are almost no light thermalized neutrino species, but of course the 
diluters' decays into neutrinos lead to nonzero effective neutrino number as discussed below.

\paragraph{CMB.}
The moment of matter-radiation equality determines the number of relativistic degrees of freedom, which can be measured by observing the CMB power spectrum and is parametrized by the effective number of neutrino species, $N_{\text{eff}}$. The determination of $N_{\text{eff}}$ has been improved over the years with the most recent best fit at $N_{\text{eff}} = 4.34^{+0.86}_{-0.88}$ (68\% CL) from WMAP-7~\cite{Komatsu:2010fb}, while another recent analysis, combining the low redshift data with the 5-year WMAP analysis, comes up with a slightly different value of $N_{\text{eff}} = 3.77 \pm 0.067$ (68\% CL)~\cite{Reid:2009nq}.

In our scenario, both diluters $N_{2,3}$ decay into neutrino-rich final states. If it were not for these decays, the light neutrinos would have been extinct. Once again, this testifies a completely different history of the early universe from the standard one. More precisely, the lighter state decays as $N_3 \to \pi^+ e^- \to e^+ e^- \nu_\mu \bar \nu_\mu \nu_e$ and the heavier $N_2 \to \pi^+ \mu^- \to e^+ e^- \nu_\mu \nu_\mu \bar \nu_\mu \nu_e \bar \nu_e $ (or their anti-particles), where the average energy of the final state neutrinos ranges from $10 - 50 \text{ MeV}$. 
As a rough estimate we simply count the number of neutrinos produced in the decay. Using Eq.~\eqref{Nyield} with $g_*(T_f) \simeq 20$, appropriate for the freeze-out of $N_{2,3}$ below $T_{\rm {QCD}}$, it implies that each of the $N_{2,3}$ number density counts as a half of the usual light neutrino one (recall that $g_*(\text {MeV}) \simeq 10$). Since on average they decay into four light neutrinos, one
gets roughly $N_{\text{eff}} \simeq 4$, however, we expect that a more precise numerical solution would give a somewhat lower value.
Namely, the weak interaction cross-section of neutrinos scattering on electrons or protons is enhanced by the neutrino energy~\cite{Strumia:2006db}, compared to the thermal one, and therefore these neutrinos down-scatter until they lose most of their energy to the plasma~\cite{Fuller:2011qy}. They will also start to annihilate with each other when becoming sufficiently populated~\cite{Hannestad:2004px}. A complete analysis of neutrino thermalization in this scenario is beyond the scope of this paper.  
\paragraph{BBN.}
A late decaying particle resulting in a low reheating temperature before the onset of BBN, could drastically change the prediction for the primordial Helium abundance by affecting the neutron-proton number ratio. 

The injection of energetic pions resulting from the $N_{2,3}$ decay would increase the neutron-proton ratio. Notice however, that the pions in this scenario are rather soft, since $m_N$ lies near the threshold, therefore they do not scatter with nucleons before decaying~\cite{Rehm:2000ai} and the above problem can easily be evaded. Second, the lack of neutrino thermalization changes both, the contribution of neutrinos to the Hubble rate, and the average weak interaction rate with competing effects. The analysis of~\cite{Kawasaki:2000en, Hannestad:2004px, Ruchayskiy:2012si} shows that a reheating temperature as low as 0.7 MeV could still be compatible with the observed Helium abundance.

\subsection{Neutrino-less double-$\beta$ decay}

As mentioned in the introduction, the Majorana nature of the RH neutrinos and the associated lepton number violation (LNV) plays a crucial role in the study of dark matter in the minimal LR model. The textbook example of LNV is the neutrinoless double beta decay ($0\nu2\beta$), often erroneously associated with neutrino mass only in spite of the fact that more than 50 years ago it was argued that new physics may be equally responsible for this process~\cite{Feinberg:1959, Pontecorvo:1968wp}.

The LR theory is tailor-made for the new physics contribution to the $0\nu2\beta$. Since left implies right, there must  of course be a right-handed counterpart to the usual left-handed neutrino contribution~\cite{Mohapatra:1980yp}. Recently, an in-depth study has been performed~\cite{Tello:2010am}, which emphasized the profound connection between $0\nu2\beta$, LNV at colliders~\cite{Keung:1983uu} and lepton flavor violation (see also~\cite{Chakrabortty:2012mh} for a recent study of $0\nu2\beta$ in LRSM and~\cite{Barry:2012ga} for a study on future linear collider signals). This leads to a derivation of a serious constraint~\cite{Nemevsek:2011aa} on the scales of LR symmetry, which depends on the flavor structure of $V^R_{\ell i}$. In our case, since $\mathbf{V_\ell^R}$ ends up being diagonal, the limit becomes quite acute. From Fig.~1 in~\cite{Nemevsek:2011aa} one can conclude that for the preferred value of $m_N$ around 140 MeV, the mass of $W_R$ should lie above $\sim$\,6\,--\,8\,TeV. Given the large uncertainties of the nuclear matrix elements~\cite{Rodejohann:2011mu, GomezCadenas:2011it}, we cannot claim with sufficient certainty that this setup is completely outside the range of the LHC and we leave the precise determination of this particular bound to the experts. In any case, upcoming experiments are about to probe this region~\cite{KamLANDZen:2012aa, Auger:2012ar}.

%
%
\section{Conclusions and Outlook \label{secConclusions}}

The final outcome of this study is quite simple and striking. The minimal left-right symmetric model can naturally provide a warm dark matter candidate in the form of a keV right-handed neutrino, with a hope of observing the RH charged gauge boson $W_R$ at the LHC. For those of you who do not care about the nitty-gritty of the theory and our analysis, this is the main message we would like you to take home.

For this to be true, the following spectrum emerges, $m_{N_1}\simeq$\,keV (Dark Matter), $m_{N_2}\simeq m_\pi+m_\mu$, $m_{N_3}\simeq m_\pi+m_\mu$ (Diluters); together with a particular flavor-diagonal structure $V^R_{\tau1} \simeq V^R_{\mu2}\simeq V^R_{e3} \simeq 1$. This is the only option to have a reasonably light $W_R$ accessible at the LHC. We find on top, for $m_{N_1}$ lying around 0.5 keV, a narrow window of $M_{W_R}$ around 5\,TeV, otherwise $M_{W_R}$ has to be larger than about 20\,TeV. We find this isolated window very interesting. 

A noteworthy fact. For this DM picture to work, with the spectrum taken as above, at no point we need to assume that the universe ever reached very high temperature. The highest temperature we talk about here is less than about GeV, a modest extrapolation of the BBN temperatures. This puts the warm DM picture of the Left-Right Theory on quite firm grounds.

The million dollar question is: how to test this DM scenario of the LRSM? Ideally, one would like to directly measure the masses and mixings of the right-handed neutrinos together with a mass of $W_R$. The only way to measure $m_N$ and $\mathbf{V}^R_{\ell}$ at the LHC is through the golden KS channel~\cite{Keung:1983uu} of two charged leptons and two jets without any missing energy. Unfortunately, this requires much larger masses $m_N \gtrsim 10 \text{ GeV}$ than the resulting spectrum above, which could manifest itself simply as the missing energy due to the extremely long-lived diluters. At this point, it seems difficult to imagine a conceivable way of directly measuring $\mathbf{V}^R_{\ell}$ and/or $m_N$, despite the fact that the resulting parameter space in the right-handed leptonic sector is well determined. On the other hand, one can search for indirect signals using low-energy processes, similarly to the sterile neutrino case~\cite{Gorbunov:2007ak, Atre:2009rg, Mitra:2011qr} (see also~\cite{deVega:2011xh, Li:2010vy, Li:2011mw, Chakrabortty:2012pp}), which have a correlated dependence on both, the mixing parameters and the mass spectrum. In view of the smallness of the Dirac Yukawa couplings, the presence of new heavy gauge bosons is more than welcome for the sake of visibility of these elusive, otherwise sterile neutrinos. Moreover, a $W_R$ in the $\sim 5 \text{ TeV}$ window can well be probed at 14 TeV LHC in the case when light RH neutrinos escape detection with a potential to establish the chirality of the outgoing fermions~\cite{Ferrari:2000sp}.~\footnote{Still, one of the authors of the paper (GS) is deeply disturbed by the sad outcome of this work, which says that the golden KS channel will not be measured at the LHC.}

On a positive note, we predict a sizable rate for the neutrino-less double beta decay, already on the edge of exclusion by the current data. It is perhaps even more difficult to imagine that $0\nu2\beta$ decay would not be seen in the currently on-going and planned experiments, if $W_R$ were seen at the LHC. In this sense, our work provides an additional impetus for the dedicated $0\nu2\beta$ decay searches. Moreover, although it is hard to verify precisely the picture that emerges from our analysis, it is easy to kill it. The smallness of RH neutrino masses and the absence of leptonic mixing, eliminates completely lepton number violation at colliders and any lepton number violation. Observing any such process would invalidate completely the scenario offered here and dark matter would need another origin.

In our opinion, there are still some issues that deserve deeper understanding and more quantitative study. The foremost is the neutrino thermalization before the BBN epoch, which affects both the light element production and the effective neutrino species $N_{\rm eff}$ measured by CMB (see e.g.~\cite{Fuller:2011qy}). This could seal the fate of our scenario, especially in view of better sensitivity of the upper coming data~\cite{Ade:2011ap, Das:2011ak, Keisler:2011aw}. Furthermore, there seems to be a lower limit on the X-ray line flux from the decay of our warm dark matter, due to the flavor structure imposed by left-right symmetry and the constraints from our dark matter analysis.

\section*{Acknowledgement}

We wish to thank Francesco Vissani for his interest in this important question and for bringing up the issue of dark matter in left-right theory some time ago. We are grateful to Alejandra Melfo and Fabrizio Nesti for careful reading of the manuscript and useful comments. We would like to thank Michele Frigerio, Julien Lavalle, Gilbert Moultaka, Alexei Smirnov and Vladimir Tello for useful discussions and comments. MN acknowledges the support of the mobility grant ``Leonardo da Vinci - Vse\v{z}ivljensko u\v{c}enje'', LDV-MOB-74/11. 

\appendix 

%
%
\section{Annihilation Cross Sections \label{AppndxXs}}

In order to write down the Boltzmann equations for the freeze-out process, we summarize the main processes governing the freeze-out. In order to get the thermal reaction rates discussed in Appendix~\ref{AppndxXs2}, one needs to know annihilation cross sections. We work with the flavor structure where $N_1$ mainly couples to $\tau$. When the temperature drops well below $m_\tau$, all charge current interactions of $N_1$ become negligible due to Boltzmann suppression.

One actually uses the so-called reduced cross section (we consider $2\to2$ processes $1,2\to a,b$ only), defined as
\begin{eqnarray}
\!\!\hat \sigma (s) \equiv \frac{2 \sqrt{(p_1\cdot p_2)^2 - m_1^2 m_2^2}}{s} \!\int\! \frac{d^3 p_a}{(2\pi)^3} \frac{1}{2E_a} \frac{d^3 p_b}{(2\pi)^3} \frac{1}{2E_b} (2\pi)^4 \delta^{(4)}(p_1+p_2-p_a-p_b) \overline{|\mathcal{M}|^2} . 
\end{eqnarray}

\begin{enumerate}[a)]

\item { \it Single-$N$ annihilations, $W_R$-exchange} \\
The single-$N$ annihilation processes are mediated by $W_R$ in both s- and t-channels.  In the case massive $N_{2,3}$, the reduced cross sections are
\begin{align} \label{sigmasingleNd}
\hat \sigma_{t}^{W_R} (Nd_R\to e_R u_R)(s) &= \frac{N_c g^4 (m_{N}^6 - 3 m_{N}^2 s^2 + 2 s^3)}{48\pi M_{W_R}^4 s} \ ,
\\
\label{sigmasingleNu}
\hat \sigma_{t}^{W_R} (N\bar u_R\to e_R \bar d_R)(s) &= \frac{N_c g^4 (m_{N}^2 -s)^2}{48\pi M_{W_R}^4} \ ,
\\
\label{sigmasingleNe}
\hat \sigma_{s}^{W_R} (N\bar e_R\to e_R \bar d_R)(s) &= \frac{N_c g^4 (m_{N}^2 -s)^2(m_{N}^2 +2s)}{48\pi M_{W_R}^4s}  \ ,
\end{align}
where $g_R$ is the $SU(2)_R$ gauge coupling, $N_c=3$ is the color factor and $s$ is the center-of-mass energy squared of the $2\to2$ scattering process. Our results agree with those calculated in Ref.~\cite{Frere:2008ct}, up to the number of generations. Since we are discussing physics with energy scale around and below GeV, the $1/M_{W_R}^2$ expansion has been made in the above expressions. In the Boltzmann equations, we also take into account of the charge conjugation processes.

In the case of $N_1$, which is effectively massless, we have to keep the mass of $\tau$ lepton to which it predominantly couples. The corresponding cross sections can be obtained by simply making the replacement $m_N\to m_\tau$ in Eqs.~\eqref{sigmasingleNd}-\eqref{sigmasingleNe}.

\item {\it Pair-of-$N$ annihilation to the same flavor charge leptons, $W_R$ and $Z_{LR}$-exchanges}\\

The pair-annihilation process can take place through t- and u-channel $W_R$ exchange, as well as s-channel $Z_{LR}$ exchange. For the case of $N_{2,3}$, the final leptons are massless and the cross section is
\begin{equation}
\begin{split}
&\hat \sigma_{t+u}^{W_R, Z_{LR}} (N N\to \ell^+ \ell^-)(s) = \frac{g^4}{12 \pi M_{W_R}^4} \sqrt{s} (s-4m_{N}^2 )^{3/2}
\\
&\ - \frac{g^4 (3 \cos2\theta_W -1) \sec^2\theta_W \sec 2\theta_W}{192 \pi M_{W_R}^2 M_{Z_{LR}}^2} \sqrt{s} (s-4m_N^2)^{1/2} \left(s- 4m_N^2 + (s+2 m_N^2) \cos2\theta_W \rule{0cm}{4mm}\right)
\\
&\ + \frac{g^4 \sec^8 \theta_W}{3072 \pi (1-\tan^2\theta_W)^2 M_{Z_{LR}}^4 s} \left[ 16 \cos^2 2\theta_W (2 \cos2\theta_W -1) \left( (s-m_N^2)^{3} - s^3 + m_N^6 \rule{0cm}{4mm}\right) \rule{0cm}{6mm}\right.
\\
& \left.\ + (8 \cos 2\theta_W - 5 \cos4\theta_W -7) (m_N^2-s) \left( \cos4\theta_W (s + 2 m_N^2) (2 s+ m_N^2) + 6 m_N^4 - 9 m_N^2 s + 6 s^2  \rule{0cm}{4mm}\right)  \rule{0cm}{6mm}\right]
\end{split}
\end{equation}
For the case of $N_1$, where the mass of $\tau$ in the final states cannot be neglected. Here the t-, u-channel interference is suppressed by the smallness of $N_1$ mass. The cross section is
\begin{equation}
\begin{split}
&\hat \sigma_{t+u}^{W_R, Z_{LR}} (N N\to \ell^+ \ell^-)(s) = \frac{g^4}{12 \pi M_{W_R}^4} \sqrt{s} (s-4m_{N}^2)^{1/2} (s - m_\tau^2 ) 
\\
&\ - \frac{g^4}{96 \pi M_{W_R}^2 M_{Z_{LR}}^2} \sqrt{s} (s-4m_N^2)^{1/2} \left(3s -6m_\tau^2 - (s - 4 m_\tau^2) \sec2\theta_W \rule{0cm}{4mm}\right)
\\
&\ + \frac{g^4 \sec^2 \theta_W (m_\tau^2 -s)}{768 \pi (1-\tan^2\theta_W)^2 M_{Z_{LR}}^4 s} \left[ 12 ( 12 \cos2\theta_W + 5 (2\cos2\theta_W + \cos4\theta_W+2) \sec^6 \theta_W  \rule{0cm}{6mm}\right. 
\\
& \left.\ -36 \rule{0cm}{4mm} ) m_\tau^2 s + (8 \cos 2\theta_W - 5 \cos4\theta_W -7) \sec^2\theta_W (1+\tan^4\theta_W) \left( 2s^2 - m_\tau^2 s+ 2 m_\tau^4 \rule{0cm}{3mm}\right)  \rule{0cm}{6mm}\right]
\end{split}
\end{equation}

\item {\it Pair-of-$N$ annihilations to other fermions, $Z_{LR}$-exchange only.} The cross section of the pair-annihilation process can also take place through s-channel $Z_{LR}$ exchange
\begin{equation}
\begin{split}
&\hat \sigma^{Z_{LR}}_{NN} (NN\to f'\bar f')(s) 
= \frac{g^4 \sec^8 \theta_W}{9136 \pi (1-\tan^2\theta_W)^2 M_{Z_{LR}}^4 s} \biggl[ 48 \cos^2 2\theta_W (8 \cos2\theta_W-1)  \rule{0cm}{6mm}
\\
& \times \left( (s-m_N^2)^3 -s^3 + m_N^6 \rule{0cm}{4mm}\right) - (32 \cos 2 \theta_W + 53 \cos4\theta_W + 63) (s-m_N^2)
\\
& \times \left( \cos4\theta_W (s + 2 m_N^2) (2 s+ m_N^2)  + 6 s^2 - 9 m_N^2 s + 6 m_N^4 \rule{0cm}{4mm}\right) \rule{0cm}{6mm}\biggr] \ ,
\end{split}
\end{equation}
where $f'$ is any fermion except for a charged lepton that couples to $N$ and $W_R$ (recall that the RH leptonic mixing is flavor diagonal, as discussed throughout the section sec.~\ref{secTwoNs}).
For the case of $N_1$, one can simply take the limit $m_{N}\to0$.

\end{enumerate}

\section{Effective Thermal Rates \label{AppndxXs2}}
 
The processes involving the inter-flavor pair annihilation of $N$'s via $W_R$ are always subdominant and have been neglected.  For $2\to 2$ processes, the reaction rate weighted by thermal distribution is
\begin{equation}
  \gamma(1,2\to a,b)\equiv \frac{T_\gamma}{64\pi^4} \int_{s_{min}}^\infty ds \hat \sigma(s) \sqrt{s} K_1\left( \frac{\sqrt{s}}{T_\gamma} \right) \ .
\end{equation}
Here, the photon temperature $T_\gamma$ is used because the thermal averaged rates are for those particles within equilibrium with the photon and $s={\rm Min}\left\{ (m_1+m_2)^2, (m_a+m_b)^2 \right\}$.

\begin{figure}[t]
\vspace{-2ex}%
\centerline{\includegraphics[width=8cm]{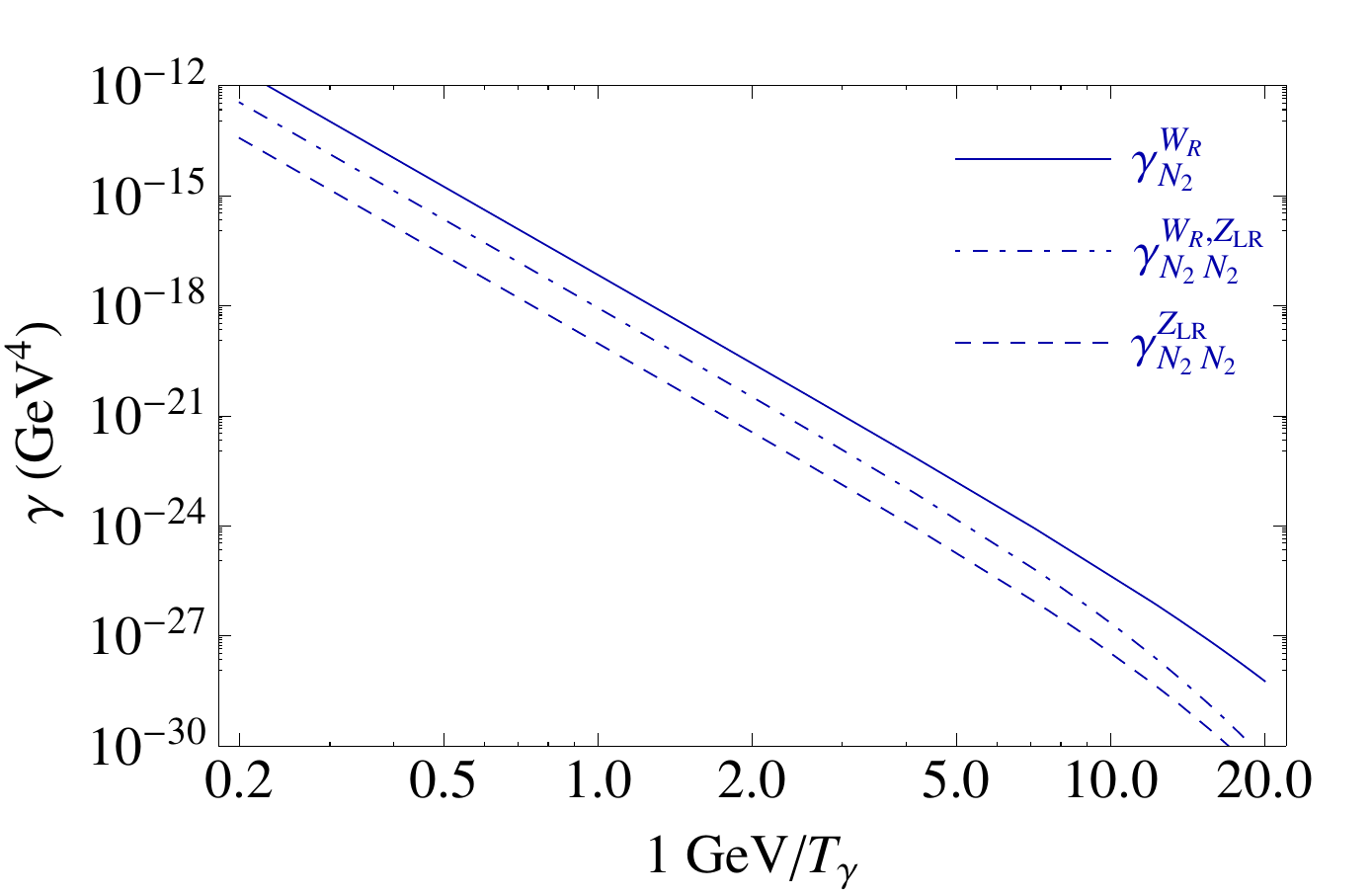}
\includegraphics[width=8cm]{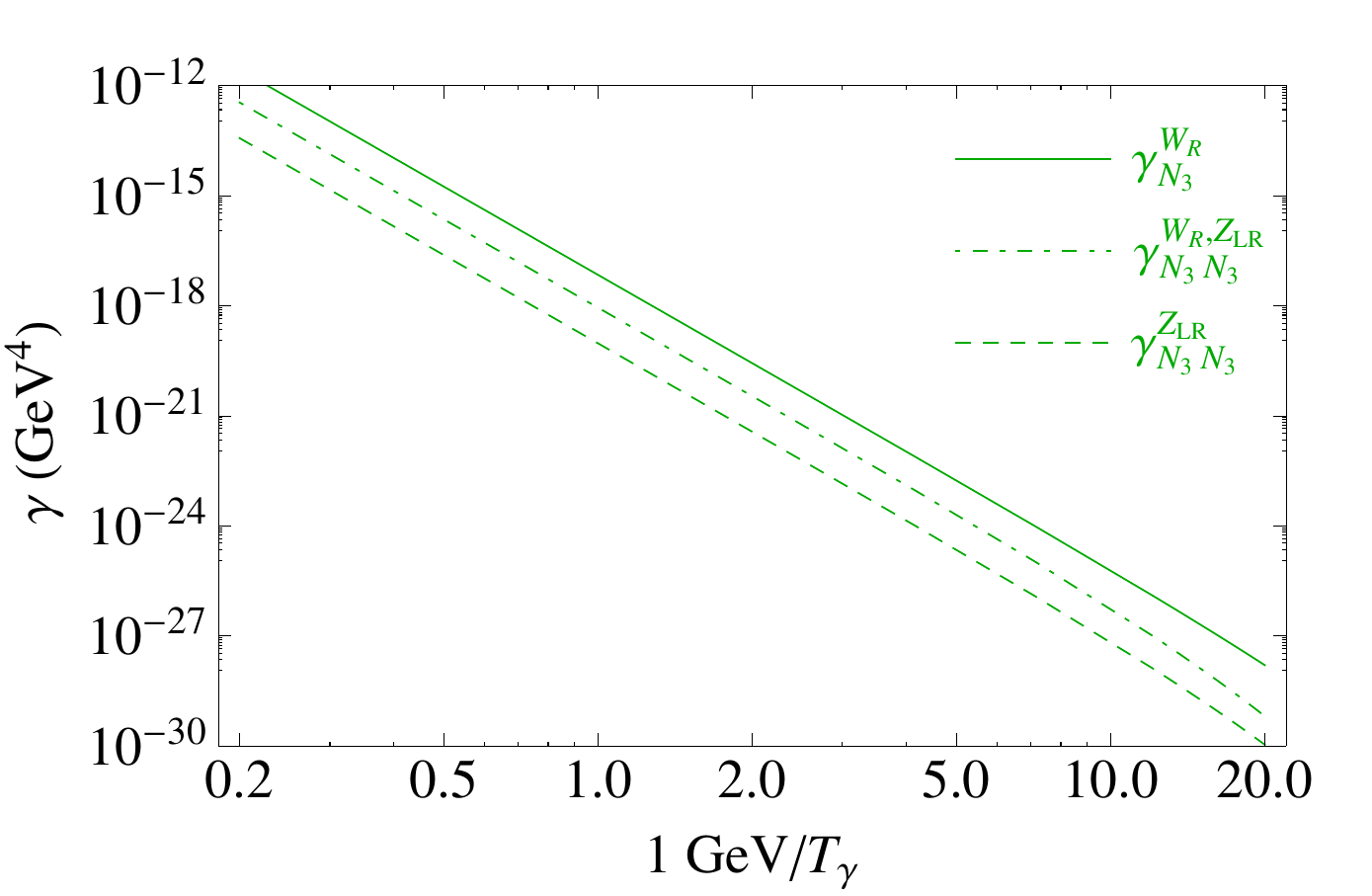}}
\centerline{\includegraphics[width=8cm]{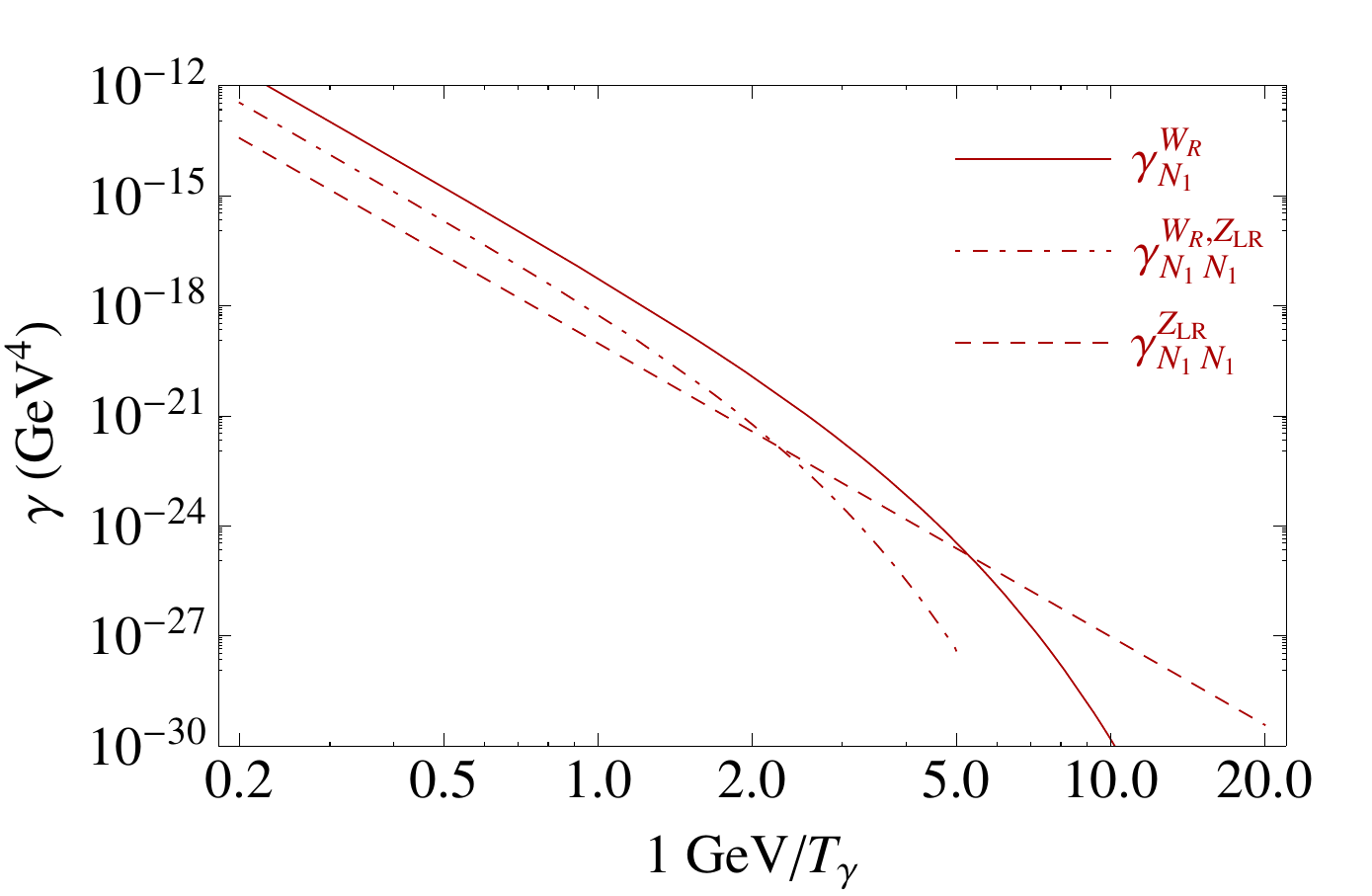}}
\vspace{-2ex}%
\caption{Thermal averaged reaction rates for the processes that controls the decoupling of $N_1$ (red), $N_2$ (blue) and $N_3$ (green), for parameters $M_{W_R}=5\,$TeV, $m_{N_2}=0.25$\,GeV, $m_{N_3}=0.14$\,GeV. For each $N$, the relative value of different rates does not depend on the choice of $M_{W_R}$.}
\label{figGamma123}
\vspace{-1.ex}
\end{figure}

We name the reactions rates $\gamma_{N_i}^{W_R}$ for the sum of processes in a), $\gamma_{N_iN_i}^{W_R, Z_{LR}}$ for process b) and $\gamma_{N_iN_i}^{Z_{LR}}$ for process c) of the Appendix~\ref{AppndxXs}.
Here, we list the thermal reaction rates for single-$N$ annihilations.
\begin{equation}
\begin{split}
\gamma_{N_3}^{W_R} &= 2\times \frac{3 g^4 m_{N_2}^4 m_{N_3}^4 \left[ K_2(z_\gamma m_{N_3}/m_{N_2}) + 5  m_{N_2} K_3(z_\gamma m_{N_3}/m_{N_2})/(z_\gamma m_{N_3}) \right]}{16 \pi^5 M_{W_R}^4 z_\gamma^4} \ ,
\\
\gamma_{N_2}^{W_R} &= \gamma_{N_3}^{W_R} (m_{N_3} \to m_{N_2}), \ \ \ \ \ \gamma_{N_1}^{W_R} = \gamma_{N_3}^{W_R} (m_{N_3} \to m_\tau), 
\end{split}
\end{equation}
where $z_\gamma= m_{N_2}/T_\gamma$ and the pre-factors 2 represent the contribution of charge conjugation processes. 

In Fig.~\ref{figGamma123}, we plot the the thermal averaged reaction rates for the processes that control the freeze out of $N_1$ and $N_2$, for $M_{W_R}=5\,$TeV, $m_{N_2}=0.25$\,GeV. For the case of $N_{2,3}$, the single-$N$ processes (with rate $\gamma_{N_{2,3}}^{W_R}$) always dominate. On the other hand, for $N_1$, the single-$N$ interaction receives the Boltzmann suppression due to the presence of $\tau$ lepton and become subdominant for temperature below 200\,MeV.

In Fig.~\ref{figGammaH}, we plot the the thermal averaged reaction rates that dominate the freeze out of $N_1$ and $N_2$, for the same set of parameters as in Fig.~\ref{figGamma123}. We also plot  the Hubble expansion rate multiplied by the thermal number density of $N_2$. Using the naive decoupling condition, $\gamma = n H$, we find the freeze out temperatures are $T_{f1}\sim 450\,$MeV and $T_{f2}\sim 250\,$MeV, respectively. This difference makes the scenario, where $N_1$ freezes out before and $N_2$ after the QCD phase transition (the tilde on the $nH$ curve), possible.

%
%

\end{document}